\documentclass[8pt]{article}
\usepackage{amsmath}
\usepackage{amssymb,scalefnt}
\usepackage{amsfonts}
\usepackage{latexsym}
\usepackage{color}
\usepackage{dsfont}
\usepackage{graphicx}
\usepackage{subfigure}
\usepackage{algorithm}
\usepackage{algorithmic}
\usepackage{pdfpages}
\usepackage{graphicx,epstopdf}
\catcode `\@=11 \@addtoreset{equation}{section}
\def\theequation{\arabic{section}.\arabic{equation}}
\catcode `\@=12
\renewcommand{\thesection}{\Roman{section}}
\newcommand{\be}{\begin{equation}}
\newcommand{\en}{\end{equation}}
\newcommand{\bea}{\begin{eqnarray}}
\newcommand{\ena}{\end{eqnarray}}
\newcommand{\beano}{\begin{eqnarray*}}
\newcommand{\enano}{\end{eqnarray*}}
\newcommand{\bee}{\begin{enumerate}}
\newcommand{\ene}{\end{enumerate}}

\newcommand{\mc}{\mathcal}

\newcommand{\Pc}{{\cal P}}
\newcommand{\Rc}{{\cal R}}
\newcommand{\Sc}{{\cal S}}

\newcommand{\F}{{\cal F}}

\newcommand{\1}{1 \!\! 1}

\newcommand{\Hil}{\mc H}

\newtheorem{thm}{Theorem}

\newtheorem{defn}[thm]{Definition}

\renewcommand{\thesection}{\arabic{section}}

\catcode `\@=11 \@addtoreset{equation}{section}
\catcode `\@=12
\begin{document}

\begin{center}
{\Large \textbf{Modeling interactions between political parties and electors}}\vspace{1.5cm}%
\\[0pt]

{\large F. Bagarello}\\[0pt]
Dipartimento di Energia, Ingegneria dell'Informazione e Modelli Matematici,\\%
[0pt]
Scuola Politecnica Ingegneria, Universit\`a di Palermo,\\[0pt]
I-90128 Palermo, Italy\\[0pt]
and I.N.F.N., Sezione di Torino\\[0pt]
e-mail: fabio.bagarello@unipa.it\\[0pt]
home page: www.unipa.it/fabio.bagarello\\[10pt]

{\large F. Gargano}\\[0pt]
Dipartimento di Energia, Ingegneria dell'Informazione e Modelli Matematici,\\%
[0pt]
Scuola Politecnica Ingegneria, Universit\`a di Palermo,\\[0pt]
I-90128 Palermo, Italy\\[0pt]
e-mail: francesco.gargano@unipa.it\\[0pt]

\end{center}

\begin{abstract}
In this paper we extend some recent results on an operatorial approach to the description of alliances between political parties interacting among themselves and with a basin of electors. In particular, we propose and compare three different models, deducing the dynamics of their related {\em decision functions}, i.e. the attitude of each  party to form or not an alliance. In the first model the interactions between each party and their electors are considered. We show that these interactions drive the decision functions towards certain asymptotic values depending on the electors only: this is the {\em perfect party}, which behaves following the electors' suggestions.  The second model is an extension of the first one in which we include a $rule$ which modifies the status of the electors, and of the decision functions as a consequence, at some specific time step. In the  third model we neglect the interactions with the electors while we consider cubic and quartic interactions between the parties and we show that  we get (slightly  oscillating) asymptotic values for the decision functions, close to their initial values. This is the {\em real party}, which does not listen to the electors. Several explicit situations are considered in details and numerical results are also shown.
\end{abstract}


\section{Introduction}

Mathematical modeling is a huge field of research which is applied to many different contexts, from physics to biology, from chemistry to finance. Tools and ideas usually adopted in physics have widely been used in these contexts. {Pars pro toto, we cite \cite{mantegna}, which is now considered a milestone in econophysics, \cite{castellano,CCC06,ST06}, where the framework of statistical physics is used outside physics, and \cite{PS07,PT14} as application of stochastic dynamics to social systems.}  Among all the strategies adopted during the years to build up models of some specific phenomenon,  in recent years many researchers started to use methods typically connected with quantum mechanics, even when dealing with macroscopic systems. This has been done in  decision-making processes, \cite{qdm1,qdm2,qdm3,qdm4,qdm5,bagbook}, in population dynamics, \cite{BGO,Gar}, in ecological processes, \cite{BO14,BCO16,DO16}, and, recently, in the analysis of political systems, \cite{pol1,pol4,all1,all2,all4}.
In these latter papers, the general operatorial settings  analyzed in details in \cite{bagbook} have  been used in the description of a {\em political  system} consisting of three parties interacting among them and with a basin of electors and of undecided voters. The focus was on the so-called {\em decision function} (DF) of each party, i.e. on the attitude of the parties to form, or not, some alliance with the other parties.
The main ingredient needed in the analysis of the time evolution of these DFs is a suitable Hamiltonian (i.e. a self-adjoint unbounded operator) which implements in itself all the mechanisms which are expected to take place in a realistic (but simplified) political system in which each party is treated as a single {\em variable}  of the model.  This is different from what was proposed in other mathematical models on the same subject, where the focus was on the single politician's  behavior, see \cite{FW94} and \cite{pol3}.

In this paper the crucial role in the procedure of decision-making of the parties is played by the various interactions existing between the parties and the basin of electors, which we treat  as a kind of \textit{open system} (the parties) interacting with a {\em reservoir} made of infinite degrees of freedom (the electors). This approach gives rise to several scenarios which are investigated all along the paper.

We also consider the possibility  of implementing some effect which cannot be easily included in the Hamiltonian $H$ of the system. In particular we adopt a general procedure proposed first in \cite{BDGO}, where the notion of  $(H,\rho)$-induced dynamics was introduced. In particular, in \cite{BDGO}, a {\em rule} $\rho$ was added to $H$ in the analysis of the quantum game of life: $\rho$ is what is used to define the new state of the system at each iteration, because, for instance, the idea of the electors could be changed by what the parties are doing. Stated in different words, $\rho$ is used to prepare the physical system for the iteration $k+1$ once the iteration $k$ is performed. This enriches quite a lot the  dynamics, and in fact several interesting results are deduced. In particular, we will discuss how a suitable rule, introduced to mimic the effect of the information {\em coming from the electors and reaching the parties}, can really deform the dynamics.

We will finally consider also the effect of  nonlinear terms in the equations of motion. Something similar was  done in \cite{all2}, where the equations of motion were solved perturbatively, while here, paying the price of neglecting the interactions of the parties with their electors (i.e. taking the parameters measuring these interactions to be zero) and focusing only on the interactions among the parties, we are able to get analytical solutions.

The paper is organized as follows: in the next section we briefly discuss the model in \cite{all4}, and we discuss what this model produces when the parameters are changed. Section \ref{sectrule} is devoted to the analysis of a deformed version of the previous model, deformation induced by the presence of a rule $\rho$. In Section \ref{sectnonlinmod} we propose a non quadratic Hamiltonian producing nonlinear, but still exactly solvable, equations of motion. Section \ref{sectconcl} contains our conclusions. To keep the paper self-consistent, the Appendix contains some essential facts on the $(H,\rho)$-induced dynamics.

\section{The first model}\label{sec:themodel}

In this section, following \cite{all1,all4}, we consider a {\em physical system} $\Sc$ consisting, first of all, of three parties, $\Pc_1$, $\Pc_2$ and $\Pc_3$, which, together, form what we call $\Sc_\Pc$. Each party has to make a choice, and it can only choose one or zero, corresponding respectively to {\em form a coalition} with some other party or not.  Hence we have $2^3=8$ different possibilities, which we associate to eight different and mutually orthogonal vectors in an eight-dimensional Hilbert space $\Hil_\Pc$. These vectors are called $\varphi_{i,k,l}$, with $i,k,l=0,1$. The three subscripts refer to whether or not the three parties of the model want to form a coalition at time $t=0$. Hence, for example, the vector $\varphi_{0,0,0}$, describes the fact that, at $t=0$, no party wants to ally with the other parties. Of course, this attitude can change during the time evolution, and deducing these changes is, in fact, what is interesting for us. {This will be achieved, see below, by considering the mean values of some particular operators on these vectors or on some of their linear combinations}.  The set $\F_\varphi=\{\varphi_{i,k,l},\,i,k,l=0,1\}$ is an orthonormal basis for $\Hil_\Pc$. In general, a vector $\Psi_0=\sum_{i,k,l}\alpha_{i,k,l}\varphi_{i,k,l}$, with $\sum_{i,k,l}|\alpha_{i,k,l}|^2=1$, can be interpreted as a vector on $\Sc_\Pc$ in which the probability of finding $\Sc_\Pc$ in a state $\varphi_{i,k,l}$, at $t=0$, is given by $|\alpha_{i,k,l}|^2$. {In particular, if for instance $\Psi_0=\varphi_{0,1,0}$, then the probability that, at $t=0$, $\Pc_1$ and $\Pc_3$ do not want to form any alliance while $\Pc_2$ does, is equal to one.}

As it is shown in \cite{all1}, it is convenient to construct the vectors $\varphi_{i,k,l}$ in a very special way, starting with the vacuum of three fermionic operators, $p_1$, $p_2$ and $p_3$, i.e. three operators which, together with their adjoints $p_1^\dagger$, $p_2^\dagger$ and $p_3^\dagger$, satisfy the canonical anticommutation relations (CAR) $\{p_k,p_l^\dagger\}=\delta_{k,l}$ and $\{p_k,p_l\}=0$. Then,  $\varphi_{0,0,0}$ is a vector satisfying $p_j\varphi_{0,0,0}=0$, $j=1,2,3$, and the other vectors $\varphi_{i,k,l}$  can be constructed
 out of $\varphi_{0,0,0}$ as follows:
$$
\varphi_{1,0,0}=p_1^\dagger\varphi_{0,0,0}, \quad \varphi_{0,1,0}=p_2^\dagger\varphi_{0,0,0}, \quad \varphi_{1,1,0}=p_1^\dagger\,p_2^\dagger\varphi_{0,0,0},\quad \varphi_{1,1,1}=p_1^\dagger\,p_2^\dagger\,p_3^\dagger\varphi_{0,0,0},
$$
and so on. Let now $\hat P_j=p_j^\dagger p_j$ be the so-called {\em number operator} of the $j$-th party. This operator satisfies  $\hat P_j\varphi_{n_1,n_2,n_3}=n_j\varphi_{n_1,n_2,n_3}$, for $j=1,2,3$, and the eigenvalues $n_j$ of these operators, zero and one, correspond to the only possible choices of the three parties at $t=0$, at least when  the state of the system at $t=0$ is one of the vectors $\varphi_{n_1,n_2,n_3}$. {More in general, if the initial state of the system is given by the vector $\Psi_0$ above, then
 \beano
  \hat P_1\Psi_0=\sum_{k,l}\alpha_{1,k,l}\varphi_{1,k,l}, \quad \hat P_2\Psi_0=\sum_{i,l}\alpha_{i,1,l}\varphi_{i,1,l}, \quad \hat P_3\Psi_0=\sum_{i,k}\alpha_{i,k,1}\varphi_{i,k,1},
  \enano

 and the initial mean values $\left<\hat P_j\right>$  of the operators $\hat P_j$ on $\Psi_0$ turn out to be
 \be\label{add3}
   \left<\hat P_1\right>=\sum_{k,l}|\alpha_{1,k,l}|^2, \quad \left<\hat P_2\right>=\sum_{i,l}|\alpha_{i,1,l}|^2, \quad \left<\hat P_3\right>=\sum_{i,k}|\alpha_{i,k,1}|^2,
   \en
 which are all real numbers between 0 and 1.
 This is the main reason why we have  used here the fermionic operators $p_j$: they produce a mean value of the number operators $\hat P_j$ which is always ({i.e., also during the time evolution; this will be clear later}) between 0 and 1, and  according to what was stated in \cite{all1}-\cite{all4}, and to the  definition given later in \eqref{add1}, we shall interpret this  value as a measure of the attitude of the party $\Pc_j$ to form an alliance or not, maximum when  it is one, and minimal when it is zero.}

\begin{figure}
\begin{center}
\hspace*{1.5cm}\begin{picture}(480,90)

\put(160,65){\thicklines\line(1,0){45}}
\put(160,85){\thicklines\line(1,0){45}}
\put(160,65){\thicklines\line(0,1){20}}
\put(205,65){\thicklines\line(0,1){20}}
\put(183,75){\makebox(0,0){$\Pc_2$}}

\put(300,35){\thicklines\line(1,0){45}}
\put(300,55){\thicklines\line(1,0){45}}
\put(300,35){\thicklines\line(0,1){20}}
\put(345,35){\thicklines\line(0,1){20}}
\put(323,45){\makebox(0,0){$\Pc_3$}}

\put(10,35){\thicklines\line(1,0){45}}
\put(10,55){\thicklines\line(1,0){45}}
\put(10,35){\thicklines\line(0,1){20}}
\put(55,35){\thicklines\line(0,1){20}}
\put(33,45){\makebox(0,0){$\Pc_1$}}

\put(10,-55){\thicklines\line(1,0){45}}
\put(10,-35){\thicklines\line(1,0){45}}
\put(10,-55){\thicklines\line(0,1){20}}
\put(55,-55){\thicklines\line(0,1){20}}
\put(33,-45){\makebox(0,0){$\Rc_1$}}

\put(140,-55){\thicklines\line(1,0){85}}
\put(140,-35){\thicklines\line(1,0){85}}
\put(140,-55){\thicklines\line(0,1){20}}
\put(225,-55){\thicklines\line(0,1){20}}
\put(183,-45){\makebox(0,0){$\Rc_2$}}

\put(300,-55){\thicklines\line(1,0){45}}
\put(300,-35){\thicklines\line(1,0){45}}
\put(300,-55){\thicklines\line(0,1){20}}
\put(345,-55){\thicklines\line(0,1){20}}
\put(323,-45){\makebox(0,0){$\Rc_3$}}

\put(140,-155){\thicklines\line(1,0){85}}
\put(140,-95){\thicklines\line(1,0){85}}
\put(140,-155){\thicklines\line(0,1){60}}
\put(225,-155){\thicklines\line(0,1){60}}
\put(183,-125){\makebox(0,0){$\Rc_{und}$}}

\put(70,44){\thicklines\vector(1,0){220}}
\put(70,44){\thicklines\vector(-1,0){3}}
\put(70,44){\thicklines\vector(3,1){80}}
\put(70,44){\thicklines\vector(-3,-1){3}}
\put(290,44){\thicklines\vector(-3,1){80}}
\put(290,44){\thicklines\vector(3,-1){3}}

\put(31,27){\thicklines\vector(0,-1){55}}
\put(31,27){\thicklines\vector(0,1){3}}
\put(322,27){\thicklines\vector(0,-1){55}}
\put(322,27){\thicklines\vector(0,1){3}}
\put(161,57){\thicklines\vector(0,-1){85}}
\put(161,57){\thicklines\vector(0,1){3}}

\put(35,27){\thicklines\vector(1,-1){115}}
\put(35,27){\thicklines\vector(-1,1){3}}
\put(318,27){\thicklines\vector(-1,-1){115}}
\put(318,27){\thicklines\vector(1,1){3}}
\put(195,57){\thicklines\vector(0,-1){145}}
\put(195,57){\thicklines\vector(0,1){3}}

\put(35,27){\thicklines\vector(2,-1){105}}
\put(35,27){\thicklines\vector(4,-1){255}}
\put(318,27){\thicklines\vector(-2,-1){105}}
\put(318,27){\thicklines\vector(-4,-1){255}}
\put(195,57){\thicklines\vector(1,-1){95}}
\put(161,58){\thicklines\vector(-1,-1){95}}


\end{picture}
 \end{center}
\vspace*{5.3cm}
\caption{\label{figscheme} The system and its multi-component reservoir.}
\end{figure}
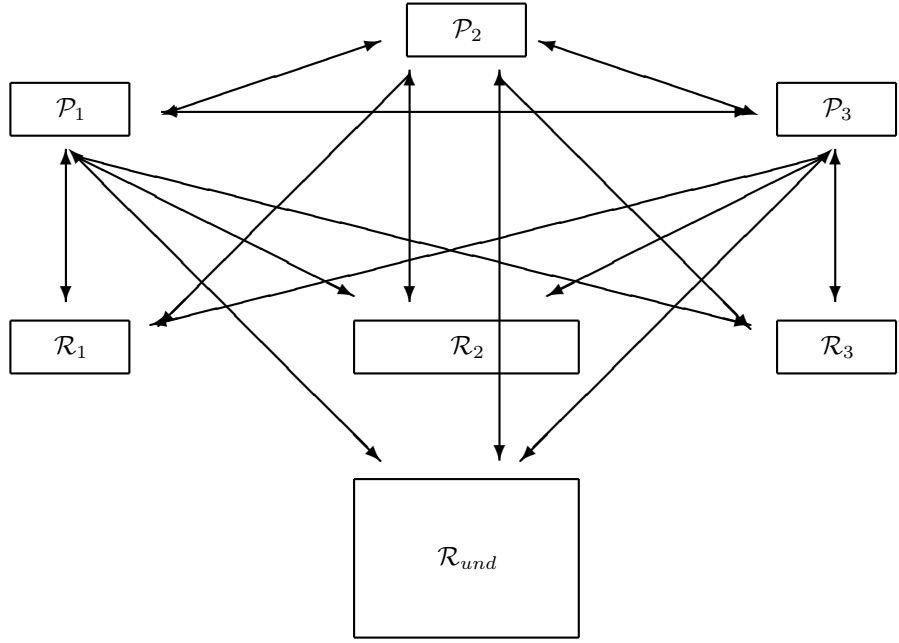

Following the general scheme described in \cite{bagbook} and adopted in very different contexts, trough the Heisenberg equation for the operators we are able to give a dynamics to the number operator $\hat P_j$, and, consequently, a time evolution to the mean values  $\left<\hat P_j\right>$ in (\ref{add3}). Of course, these values depend on the values of the $\alpha_{i,k,l}$ at $t=0$. This will be evident in the examples discussed all along this paper. In this way, we can follow how the parties modify their attitude on forming alliances with time. This is achieved by fixing, first of all, a suitable Hamiltonian, which describes the interactions indicated by the arrows in Figure \ref{figscheme}. Here $\Rc_j$ represents the set of the supporters of $\Pc_j$, while $\Rc_{und}$ is the set of all the undecided electors. The full system $\Sc$ is the union of $\Sc_\Pc$ and $\Rc=\{\Rc_1,\Rc_2,\Rc_3,\Rc_{und}\}$. Figure \ref{figscheme} shows that, as it is reasonable to imagine, each party $\Pc_j$ interacts with the other parties, with their electors, $\Rc_j$, with the undecided voters $\Rc_{und}$, but also with the electors of the other parties, $\Rc_k$, with $k\neq j$. This possibility, in particular, was not considered in \cite{all1} and was first introduced, but not analyzed in details, in \cite{all4}.

 \vspace*{1cm}

Following what done in \cite{all1} and \cite{all4}, we assume that the Hamiltonian which describes the scheme in Figure \ref{figscheme},  written in terms of fermionic operators, is the following:

\be
\left\{
\begin{array}{ll}
&h=H+H_{mix},   \\
&H_{mix}=H_{mix}^p+H_{mix}^{ap},   \\
&H_{mix}^p=\sum_{n\neq l,1}^3\nu_{nl}^p\int_{\Bbb R}dk\left(p_nB_l^\dagger(k)+B_l(k)p_n^\dagger\right),\\
&H_{mix}^{ap}=\sum_{n\neq l,1}^3\nu_{nl}^{ap}\int_{\Bbb R}dk\left(p_n^\dagger B_l^\dagger(k)+B_l(k)p_n\right),\\
\end{array}%
\right.
\label{31}\en

where

\be
\left\{
\begin{array}{ll}
H=H_{0}+H_{PBs}+H_{PB}+H_{int}, &  \\
H_{0}=\sum_{j=1}^{3}\omega _{j}p_j^\dagger p_j+\sum_{j=1}^{3}\int_{\Bbb R}\Omega_j(k)B_j^\dagger(k)B_j(k)\,dk+\int_{\Bbb R}\Omega(k)B^\dagger(k)B(k)\,dk,   \\
H_{PBs}=\sum_{j=1}^{3}\lambda_j\int_{\Bbb R}\left(p_j B_j^\dagger(k)+B_j(k)p_j^\dagger\right)\,dk,\\
H_{PB}=\sum_{j=1}^{3}\tilde\lambda_j\int_{\Bbb R}\left(p_j B^\dagger(k)+B(k)p_j^\dagger\right)\,dk,\\
H_{int}=\mu_{12}^{ex}\left(p_1^\dagger p_2+p_2^\dagger p_1\right)+\mu_{12}^{coop}\left(p_1^\dagger p_2^\dagger+p_2 p_1\right)+\mu_{13}^{ex}\left(p_1^\dagger p_3+p_3^\dagger p_1\right)+   \\
\qquad +\mu_{13}^{coop}\left(p_1^\dagger p_3^\dagger+p_3 p_1\right)+\mu_{23}^{ex}\left(p_2^\dagger p_3+p_3^\dagger p_2\right)+\mu_{23}^{coop}\left(p_2^\dagger p_3^\dagger+p_3 p_2\right).
\end{array}%
\right.
\label{22}\en

Here $\omega _{j}$, $\lambda_j$, $\tilde\lambda_j$, $\mu_{ij}^{ex}$, $\mu_{ij}^{coop}$, $\nu_{nl}^{p}$ and $\nu_{nl}^{ap}$ are real quantities, while $\Omega_j(k)=\Omega_j k$ and $\Omega(k)=\Omega k$ are real-valued functions, $\Omega_j,\Omega>0$, whose meaning is explained in details in \cite{all1}-\cite{all4}, together with the meaning of each term of $h$.  Here we just recall that the parameters $\omega_j$,  $\Omega_j$ and $\Omega$ in $H_0$ are related to a sort of {\em inertia} of the actors of the system, see also \cite{bagbook} for a general discussion on this aspect.
The term $H_{int}$  is an interaction part of the Hamiltonian, whose parameters $\mu_{jk}^{ex}$ and $\mu_{jk}^{coop}$ measure respectively the strength of the interactions between $\Pc_j$ and $\Pc_k$; {  we observe that the interaction governed by the term $\mu_{kj}^{ex}\left(p_k^\dagger p_j+p_j^\dagger p_k\right)$ introduces some kind of competition between the parties $\Pc_k$ and $\Pc_j$, because $\mu_{kj}^{ex}\,p_k^\dagger p_j$ increases the attitude of $\Pc_k$ to ally whereas the attitude of $\Pc_j$ decreases, while the hermitian conjugate term\footnote{This kind of terms are needed to keep the Hamiltonian Hermitian.} induces an opposite effect. On the other hand, in $\mu_{kj}^{coop}\left(p_k^\dagger p_j^\dagger+p_j p_k\right)$, the two parties act in a similar way: they both increase, or decrease, their tendency to form some alliance.}
 The term $H_{PBs}$  describes the interactions between the parties and their electors, whereas $H_{PB}$ describes the interactions between the parties and the undecided electors: both these contributions model the fact that the value of the DFs are somehow driven to the input of the electors, in a rather direct way. It is important to stress  that, as it is clear from (\ref{31}) and (\ref{22}), the three parties are considered here as a part of a larger system: in order to take their decisions, the parties need to interact with the electors and among themselves, since it is exactly this interaction which motivates their final decisions. This is why $\Sc_\Pc$ must be {\em open}, i.e. there must be some environment, $\Rc$ (the full set of electors),  interacting with $\Pc_1$, $\Pc_2$ and $\Pc_3$, which produces some  feedback used by $\Pc_j$ to decide what to do. Moreover, the environment, when compared with $\Sc_\Pc$, is expected to be very large, since the sets of the electors for $\Pc_1$, $\Pc_2$ and $\Pc_3$ are supposed to be sufficiently large. This is the reason why (infinitely many) operators $B_j(k)$ and $B_j^\dagger(k)$, $k\in\Bbb R$, appear in (\ref{31}) and (\ref{22}). It can be useful to observe that $h$ is quadratic. This is physically meaningful and technically useful, since the differential equations we deduce out of $h$ can be solved analytically. We refer to \cite{bagbook} for examples, mainly in finance, where only perturbative approaches can be adopted to solve differential equations arising from, say, cubic or quartic Hamiltonians. 

 The following CAR's for the operators of the reservoir are assumed:
\be
\{B_j(k),B_l^\dagger(q)\}=\delta_{j,l}\delta(k-q)\,\1,\qquad \{B_j(k),B_l(k)\}=0,
\label{23}
\en
as well as
\be
\{B(k),B^\dagger(q)\}=\delta(k-q)\,\1,\quad \{B(k),B(k)\}=0,
\label{23b}
\en
for all $j,l=1,2,3$, $k,\, q\in{\Bbb R}$. Moreover each $p_j^\sharp$ anti-commutes with each $B_l^\sharp(k)$ and with $B^\sharp(k)$: $\{p_j^\sharp, B_l^\sharp(k)\}=\{p_j^\sharp, B^\sharp(k)\}=0$ for all $j$, $l$ and for all $k$, and we further assume that $\{B^\sharp(q), B_l^\sharp(k)\}=0$, for all $k,q\in\Bbb R$. Here $X^\sharp$ stands for $X$ or $X^\dagger$. {Of course, other choices are also possible for these operators. For instance, one could require that they obey canonical commutation relations, rather than the CAR's. However, we prefer to restrict to  (\ref{23}) and (\ref{23b}) since it reflects, for the electors, the analogous choice adopted for the operators of the parties. In this way our model is {\em entirely fermionic}. }

Once $h$ is given, we have to
compute the time evolution of the number operators in the Heisenberg scheme as $\hat P_j(t):=e^{iht}\hat P_j e^{-iht}$, and then their mean values on some suitable state describing the full system $\Sc$ (parties and electors) at $t=0$. This is what  has been called {\em decision function} (DF) in \cite{all1,all2,all4}, see also formula (\ref{add1}) below. To find these mean values it is convenient to compute first the Heisenberg equations of motion, $\dot X(t)=i[h,X(t)]$, \cite{mer,rom}, for the annihilation operators of the full system:

\be
\left\{
\begin{array}{ll}
\dot p_1(t)=-i\omega_1 p_1(t)+i\lambda_1\int_{\Bbb R}B_1(q,t)\,dq+i\tilde\lambda_1\int_{\Bbb R}B(q,t)\,dq-i\mu_{12}^{ex}p_2(t)-i\mu_{12}^{coop}p_2^\dagger(t)+\\
\qquad -i\mu_{13}^{ex}p_3(t)-i\mu_{13}^{coop}p_3^\dagger(t)+M_1(t),   \\
\vspace{1mm}
\dot p_2(t)=-i\omega_2 p_2(t)+i\lambda_2\int_{\Bbb R}B_2(q,t)\,dq+i\tilde\lambda_2\int_{\Bbb R}B(q,t)\,dq-i\mu_{12}^{ex}p_1(t)+i\mu_{12}^{coop}p_1^\dagger(t)+\\
\qquad -i\mu_{23}^{ex}p_3(t)-i\mu_{23}^{coop}p_3^\dagger(t)+M_2(t),   \\
\vspace{1mm}
\dot p_3(t)=-i\omega_3 p_3(t)+i\lambda_3\int_{\Bbb R}B_3(q,t)\,dq+i\tilde\lambda_3\int_{\Bbb R}B(q,t)\,dq-i\mu_{13}^{ex}p_1(t)+i\mu_{13}^{coop}p_1^\dagger(t)+\\
\qquad -i\mu_{23}^{ex}p_2(t)+i\mu_{23}^{coop}p_2^\dagger(t)+M_3(t),   \\
\vspace{1mm}
\dot B_j(q,t)=-i\Omega_j(q) B_j(q,t)+i\lambda_j p_j(t)+i\,R_j(t),\qquad\qquad j=1,2,3,   \\
\vspace{1mm}
\dot B(q,t)=-i\Omega(q) B(q,t)+i\sum_{j=1}^3\tilde\lambda_j p_j(t).   \label{32}
\end{array}%
\right.
\en

where we have introduced the following quantities:

\be
\left\{
\begin{array}{ll}
M_1(t)=i{\int}_{\Bbb R}\left(\nu_{12}^pB_2(q,t)+\nu_{13}^pB_3(q,t)-\nu_{12}^{ap}B_2^\dagger(q,t)-
\nu_{13}^{ap}B_3^\dagger(q,t)\right)dq\\
\vspace{1mm}
M_2(t)=i{\int}_{\Bbb R}\left(\nu_{21}^pB_1(q,t)+\nu_{23}^pB_3(q,t)-\nu_{21}^{ap}B_1^\dagger(q,t)-
\nu_{23}^{ap}B_3^\dagger(q,t)\right)dq\\
\vspace{1mm}
M_3(t)=i{\int}_{\Bbb R}\left(\nu_{31}^pB_1(q,t)+\nu_{32}^pB_2(q,t)-\nu_{31}^{ap}B_1^\dagger(q,t)-
\nu_{32}^{ap}B_2^\dagger(q,t)\right)dq\\
\vspace{1mm}
R_1(t)=\nu_{21}^pp_2(t)+\nu_{21}^{ap}p_2^\dagger(t)+\nu_{31}^pp_3(t)+\nu_{31}^{ap}p_3^\dagger(t)\\
\vspace{1mm}
R_2(t)=\nu_{12}^pp_1(t)+\nu_{12}^{ap}p_1^\dagger(t)+\nu_{32}^pp_3(t)+\nu_{32}^{ap}p_3^\dagger(t)\\
\vspace{1mm}
R_3(t)=\nu_{13}^pp_1(t)+\nu_{13}^{ap}p_1^\dagger(t)+\nu_{23}^pp_2(t)+\nu_{23}^{ap}p_2^\dagger(t).   \label{33}
\end{array}%
\right.
\en

The system in (\ref{32}) is linear in its dynamical variables, so that an analytic solution can be  found. In fact, see \cite{all4}, we deduce
\be
P(t)=e^{U\,t}P(0)+\int_0^t e^{U\,(t-t_1)}\,\eta(t_1)\,dt_1,
\label{34}\en
where we have introduced the vectors
$$
P(t)=\left(
       \begin{array}{c}
         p_1(t) \\
         p_2(t) \\
         p_3(t) \\
         p_1^\dagger(t) \\
         p_2^\dagger(t) \\
         p_3^\dagger(t) \\
       \end{array}
     \right), \quad \eta(t)=\left(
                               \begin{array}{c}
                                 \eta_1(t) \\
                                 \eta_2(t) \\
                                 \eta_3(t) \\
                                 \eta_1^\dagger(t) \\
                                 \eta_2^\dagger(t) \\
                                 \eta_3^\dagger(t) \\
                               \end{array}
                             \right),$$ and the matrix $$U=\left(
\begin{array}{cccccc}
 x_{1,1} & x_{1,2} & x_{1,3}  & y_{1,1} & y_{1,2} & y_{1,3} \\
 x_{1,2} & x_{2,2} & x_{2,3} & \overline{y_{1,2}} & y_{2,2} & y_{2,3} \\
x_{1,3} & x_{2,3} &  x_{3,3} & \overline{y_{1,3}} & \overline{y_{2,3}} &  y_{3,3} \\
 \overline{y_{1,1}} & \overline{y_{1,2}} & \overline{y_{1,3}}  & \overline{x_{1,1}} & \overline{x_{1,2}} & \overline{x_{1,3}} \\
 y_{1,2} & \overline{y_{2,2}} & \overline{y_{2,3}} & \overline{x_{1,2}} & \overline{x_{2,2}} & \overline{x_{2,3}} \\
  y_{1,3} & y_{2,3} & \overline{y_{3,3}} & \overline{x_{1,3}} & \overline{x_{2,3}} & \overline{x_{3,3}} \\
\end{array}
\right).
$$
Also,
$$
\left\{
\begin{array}{ll}
x_{1,1}=-i\omega_1-\frac{\pi}{\Omega}\,\tilde\lambda_1^2-\frac{\pi}{\Omega_1}\,\lambda_1^2-\frac{\pi}{\Omega_2}\left((\nu_{12}^p)^2+(\nu_{12}^{ap})^2\right)
-\frac{\pi}{\Omega_3}\left((\nu_{13}^p)^2+(\nu_{13}^{ap})^2\right)\\
\vspace{1mm}
x_{1,2}=-i\mu_{12}^{ex}-\frac{\pi}{\Omega}\,\tilde\lambda_1\tilde\lambda_2-\frac{\pi}{\Omega_1}\,\lambda_1\nu_{21}^p-\frac{\pi}{\Omega_2}\,\lambda_2
\nu_{12}^p-\frac{\pi}{\Omega_3}\left(\nu_{13}^p\nu_{23}^p+\nu_{13}^{ap}\nu_{23}^{ap}\right)\\
\vspace{1mm}
x_{1,3}=-i\mu_{13}^{ex}-\frac{\pi}{\Omega}\,\tilde\lambda_1\tilde\lambda_3-\frac{\pi}{\Omega_1}\,\lambda_1\nu_{31}^p
-\frac{\pi}{\Omega_2}\left(\nu_{12}^p\nu_{32}^p+\nu_{12}^{ap}\nu_{32}^{ap}\right)-\frac{\pi}{\Omega_3}\,\lambda_3
\nu_{13}^p\\
\vspace{1mm}
x_{2,2}=-i\omega_2-\frac{\pi}{\Omega}\,\tilde\lambda_2^2-\frac{\pi}{\Omega_1}\left((\nu_{21}^p)^2+(\nu_{21}^{ap})^2\right)-\frac{\pi}{\Omega_2}\,\lambda_2^2
-\frac{\pi}{\Omega_3}\left((\nu_{23}^p)^2+(\nu_{23}^{ap})^2\right)\\
\vspace{1mm}
x_{2,3}=-i\mu_{23}^{ex}-\frac{\pi}{\Omega}\,\tilde\lambda_2\tilde\lambda_3-\frac{\pi}{\Omega_1}\left(\nu_{21}^p\nu_{31}^p+\nu_{21}^{ap}\nu_{31}^{ap}\right)
-\frac{\pi}{\Omega_2}\,\lambda_2\nu_{32}^p-\frac{\pi}{\Omega_3}\,\lambda_3
\nu_{23}^p\\
\vspace{1mm}
x_{3,3}=-i\omega_3-\frac{\pi}{\Omega}\,\tilde\lambda_3^2-\frac{\pi}{\Omega_1}\left((\nu_{31}^p)^2+(\nu_{31}^{ap})^2\right)
-\frac{\pi}{\Omega_2}\left((\nu_{32}^p)^2+(\nu_{32}^{ap})^2\right)-\frac{\pi}{\Omega_3}\,\lambda_3^2,
\end{array}%
\right.
$$
\vspace{2mm}
$$
\left\{
\begin{array}{ll}
y_{1,1}=-\frac{2\pi}{\Omega_2}\,\nu_{12}^p\nu_{12}^{ap}-\frac{2\pi}{\Omega_3}\,\nu_{13}^p\nu_{13}^{ap}\\
\vspace{1mm}
y_{1,2}=-i\mu_{12}^{coop}-\frac{\pi}{\Omega_1}\,\lambda_1\nu_{21}^{ap}-\frac{\pi}{\Omega_2}\,\lambda_2\nu_{12}^{ap}-\frac{\pi}{\Omega_3}\left(
\nu_{13}^{p}\nu_{23}^{ap}+\nu_{13}^{ap}\nu_{23}^{p}\right)\\
\vspace{1mm}
y_{1,3}=-i\mu_{13}^{coop}-\frac{\pi}{\Omega_1}\,\lambda_1\nu_{31}^{ap}-\frac{\pi}{\Omega_2}\left(
\nu_{12}^{p}\nu_{32}^{ap}+\nu_{12}^{ap}\nu_{32}^{p}\right)-\frac{\pi}{\Omega_3}\lambda_3\nu_{13}^{ap}\\
\vspace{1mm}
y_{2,2}=-\frac{2\pi}{\Omega_1}\,\nu_{21}^p\nu_{21}^{ap}-\frac{2\pi}{\Omega_3}\,\nu_{23}^p\nu_{23}^{ap}\\
\vspace{1mm}
y_{2,3}=-i\mu_{23}^{coop}-\frac{\pi}{\Omega_1}\left(
\nu_{21}^{p}\nu_{31}^{ap}+\nu_{21}^{ap}\nu_{31}^{p}\right)-\frac{\pi}{\Omega_2}\lambda_2\nu_{32}^{ap}-\frac{\pi}{\Omega_3}\lambda_3\nu_{23}^{ap}\\
\vspace{1mm}
y_{3,3}=-\frac{2\pi}{\Omega_1}\,\nu_{31}^p\nu_{31}^{ap}-\frac{2\pi}{\Omega_2}\,\nu_{32}^p\nu_{32}^{ap},
\end{array}%
\right.
$$
and
$$
\left\{
\begin{array}{ll}
\eta_1(t)=i\tilde\lambda_1\beta(t)+i\lambda_1\beta_1(t)+i\nu_{12}^p\beta_2(t)+i\nu_{13}^p\beta_3(t)-i\nu_{12}^{ap}\beta_2^\dagger(t)-i\nu_{13}^{ap}\beta_3^\dagger(t)\\
\vspace{1mm}
\eta_2(t)=i\tilde\lambda_2\beta(t)+i\lambda_2\beta_2(t)+i\nu_{21}^p\beta_1(t)+i\nu_{23}^p\beta_3(t)-i\nu_{21}^{ap}\beta_1^\dagger(t)-i\nu_{23}^{ap}\beta_3^\dagger(t)\\
\vspace{1mm}
\eta_3(t)=i\tilde\lambda_3\beta(t)+i\lambda_3\beta_3(t)+i\nu_{31}^p\beta_1(t)+i\nu_{32}^p\beta_2(t)-i\nu_{31}^{ap}\beta_1^\dagger(t)-i\nu_{32}^{ap}\beta_2^\dagger(t).
\end{array}%
\right.
$$
In these last equations we have further introduced the operators $$\beta(t)=\int_{\Bbb R}B(q)e^{-i\Omega q t}dq,\qquad\beta_j(t)=\int_{\Bbb R}B_j(q)e^{-i\Omega_j q t}dq,$$ $j=1,2,3.$ Now, the time evolution of the number operators $\hat P_j(t)=p_j^\dagger(t)p_j(t)$, $j=1,2,3$, can be found, and the DFs are obtained as

\be P_j(t):=\left<\hat P_j(t)\right>=\left<p_j^\dagger(t)p_j(t)\right>,\label{add1}\en $j=1,2,3$.
 Here $\left<.\right>$ is a state over the full system. These states, \cite{bagbook}, are taken to be suitable tensor products of vector states on $\Sc_\Pc$ and states on the reservoir which obey some standard rules:  for each operator of the form $X_{\Sc}\otimes Y_{\Rc}$, $X_{\Sc}$ being an operator of $\Sc_\Pc$ and $Y_{\Rc}$ an operator of
the reservoir, i.e. any operator only refereeing to the electors, we put
\be
\left\langle X_{\Sc}\otimes Y_{\Rc}\right\rangle :=\left\langle \varphi_{n_1,n_2,n_3},X_{\Sc}\varphi_{n_1,n_2,n_3}\right\rangle \,\omega
_{\Rc}(Y_{\Rc}).
\label{add2}\en
Here $\varphi_{n_1,n_2,n_3}$ is one of the vectors introduced at the beginning of this section\footnote{More in general, we could be interested in considering a state $\left\langle X_{\Sc}\otimes Y_{\Rc}\right\rangle=\left\langle \Psi_0,X_{\Sc}\Psi_0\right\rangle \,\omega
_{\Rc}(Y_{\Rc})$, where $\Psi_0$ has also been introduced before.}, and each $n_j$ represents, as discussed before, the tendency of $\Pc_j$ to form or not some coalition at $t=0$. The state on $\Rc$ $\omega _{\Rc}(.)$ in (\ref{add2}) satisfies
the following standard properties, \cite{bagbook}:
\be
\omega _{\Rc}(1\!\!1_{\Rc})=1,\quad \omega _{\Rc}(B_{j}(k))=\omega
_{\Rc}(B_{j}^{\dagger }(k))=0,\quad \omega _{\Rc}(B_{j}^{\dagger
}(k)B_{l}(q))=N_{j}(k)\,\delta _{j,l}\delta (k-q),
\label{211}\en
as well as
\be
\omega _{\Rc}(B(k))=\omega
_{\Rc}(B^{\dagger }(k))=0,\quad \omega _{\Rc}(B^{\dagger
}(k)B(q))=N(k)\,\delta (k-q),
\label{211bis}\en
for some suitable functions $N_{j}(k)$, $N(k)$ which we take here to be constant in $k$: $N_{j}(k)=N_j$ and $N(k)=N$.  Also, we assume that $\omega
_{\Rc}(B_{j}(k)B_{l}(q))=\omega
_{\Rc}(B(k)B(q))=0$, for all $j, l=1,2,3$, and for all $k, q\in\Bbb R$. In our framework, the state in (\ref{add2}) describes the fact that, at $t=0$,  $\Pc_j$'s attitude (concerning alliances) is $n_j$ ($P_j(0)=n_j$), while the overall feeling of the voters $\Rc_j$ is $N_j$, and that of the undecided ones is $N$. {Of course, taking $N_j(k)$ and $N(k)$ to be constant might appear an oversimplifying assumption, and in fact this is correct. However, it still produces, in many concrete applications, a rather interesting dynamics for the model. This will be evident in the rest of this section.}

\vspace{2mm}

Let us now call $V_t=e^{Ut}$, which is simply the exponential of a six-by-six matrix, and let us call $(V_t)_{j,l}$ its $(j,l)$-th matrix element. Then the DFs  in (\ref{add1}) turn out to be the following functions:
\be
P_j(t)=P_j^X(t)+P_j^Y(t),
\label{35}\en
where
\be
P_j^X(t)=\sum_{l=1}^3\left((V_t)_{3+j,l}(V_t)_{j,3+l}(1-n_l)+(V_t)_{3+j,3+l}(V_t)_{j,l}n_l\right),
\label{36}\en
and
$$
P_j^Y(t)=2\pi\sum_{k,l=1}^3\int_{\Bbb R}dt_1[(V_{t-t_1})_{3+j,k}(V_{t-t_1})_{j,l}q_{k,l}^{(1)}+(V_{t-t_1})_{3+j,k}(V_{t-t_1})_{j,3+l}q_{k,l}^{(2)}+$$
\be
+(V_{t-t_1})_{3+j,3+k}(V_{t-t_1})_{j,l}q_{k,l}^{(3)}+(V_{t-t_1})_{3+j,3+k}(V_{t-t_1})_{j,3+l}q_{k,l}^{(4)}],
\label{37}\en
for $j=1,2,3$. Here, to keep the notation simple, we have introduced the following quantities:

$$
\left\{
\begin{array}{ll}
q_{1,1}^{(1)}=\frac{\nu_{12}^p\nu_{12}^{ap}}{\Omega_2}+\frac{\nu_{13}^p\nu_{13}^{ap}}{\Omega_3}\\
\vspace{1mm}
q_{1,2}^{(1)}=\frac{1}{\Omega_1}\lambda_1 \nu_{21}^{ap}(1-N_1)+\frac{1}{\Omega_2}\lambda_2 \nu_{12}^{ap}N_2+\frac{1}{\Omega_3}\left(\nu_{13}^{p}\nu_{23}^{ap}(1-N_3)+\nu_{13}^{ap}\nu_{23}^{p}N_3\right)\\
\vspace{1mm}
q_{1,3}^{(1)}=\frac{1}{\Omega_1}\lambda_1 \nu_{31}^{ap}(1-N_1)+\frac{1}{\Omega_2}\left(\nu_{12}^{p}\nu_{32}^{ap}(1-N_2)+\nu_{12}^{ap}\nu_{32}^{p}N_2\right)+
\frac{1}{\Omega_3}\lambda_3 \nu_{13}^{ap}N_3\\
\vspace{1mm}
q_{2,1}^{(1)}=\frac{1}{\Omega_1}\lambda_1 \nu_{21}^{ap}N_1+\frac{1}{\Omega_2}\lambda_2 \nu_{12}^{ap}(1-N_2)+\frac{1}{\Omega_3}\left(\nu_{13}^{p}\nu_{23}^{ap}N_3+\nu_{13}^{ap}\nu_{23}^{p}(1-N_3)\right)\\
\vspace{1mm}
q_{2,2}^{(1)}=\frac{\nu_{21}^p\nu_{21}^{ap}}{\Omega_1}+\frac{\nu_{23}^p\nu_{23}^{ap}}{\Omega_3}\\
\vspace{1mm}
q_{2,3}^{(1)}=\frac{1}{\Omega_1}\left(\nu_{21}^{p}\nu_{31}^{ap}(1-N_1)+\nu_{21}^{ap}\nu_{31}^{p}N_1\right)+\frac{1}{\Omega_2}\lambda_2 \nu_{32}^{ap}(1-N_2)+\frac{1}{\Omega_3}\lambda_3 \nu_{23}^{ap}N_3\\
\vspace{1mm}
q_{3,1}^{(1)}=\frac{1}{\Omega_1}\lambda_1 \nu_{31}^{ap}N_1+\frac{1}{\Omega_2}\left(\nu_{12}^{p}\nu_{32}^{ap}N_2+\nu_{12}^{ap}\nu_{32}^{p}(1-N_2)\right)+
\frac{1}{\Omega_3}\lambda_3 \nu_{13}^{ap}(1-N_3)\\
\vspace{1mm}
q_{3,2}^{(1)}=\frac{1}{\Omega_1}\left(\nu_{21}^{p}\nu_{31}^{ap}N_1+\nu_{21}^{ap}\nu_{31}^{p}(1-N_1)\right)+\frac{1}{\Omega_2}\lambda_2 \nu_{32}^{ap}N_2+\frac{1}{\Omega_3}\lambda_3 \nu_{23}^{ap}(1-N_3)\\
\vspace{1mm}
q_{3,3}^{(1)}=\frac{\nu_{31}^p\nu_{31}^{ap}}{\Omega_1}+\frac{\nu_{32}^p\nu_{32}^{ap}}{\Omega_2},\\
\end{array}%
\right.
$$
and
$$
\left\{
\begin{array}{ll}
q_{1,1}^{(2)}=\frac{1}{\Omega}\tilde\lambda_1^2(1-N)+\frac{1}{\Omega_1}\lambda_1^2(1-N_1)+\frac{1}{\Omega_2}\left((\nu_{12}^p)^2(1-N_2)+(\nu_{12}^{ap})^2N_2\right)+\\\vspace{2mm}
\qquad+\frac{1}{\Omega_3}\left((\nu_{13}^p)^2(1-N_3)+(\nu_{13}^{ap})^2N_3\right)\\
\vspace{1mm}
q_{1,2}^{(2)}=q_{2,1}^{(2)}=\frac{1}{\Omega}\tilde\lambda_1\tilde\lambda_2(1-N)+\frac{1}{\Omega_1}\lambda_1\nu_{21}^p(1-N_1)+\frac{1}{\Omega_2}\lambda_2
\nu_{12}^p(1-N_2)+\\\vspace{2mm}
\qquad+\frac{1}{\Omega_3}\left(\nu_{13}^p\nu_{23}^p(1-N_3)+\nu_{13}^{ap}\nu_{23}^{ap}N_3\right)\\
\vspace{1mm}
q_{1,3}^{(2)}=q_{3,1}^{(2)}=\frac{1}{\Omega}\tilde\lambda_1\tilde\lambda_3(1-N)+\frac{1}{\Omega_1}\lambda_1\nu_{31}^p(1-N_1)+
\frac{1}{\Omega_2}\left(\nu_{12}^p\nu_{32}^p(1-N_2)+\nu_{12}^{ap}\nu_{32}^{ap}N_2\right)+\\
\qquad+\frac{1}{\Omega_3}\lambda_3\nu_{13}^p\nu_{23}^p(1-N_3)\\\vspace{2mm}
\vspace{1mm}
q_{2,2}^{(2)}=\frac{1}{\Omega}\tilde\lambda_2^2(1-N)+\frac{1}{\Omega_1}\left((\nu_{21}^p)^2(1-N_1)+(\nu_{21}^{ap})^2N_1\right)+\frac{1}{\Omega_2}\lambda_2^2(1-N_2)+\\\vspace{2mm}
\qquad+\frac{1}{\Omega_3}\left((\nu_{23}^p)^2(1-N_3)+(\nu_{23}^{ap})^2N_3\right)\\
\vspace{1mm}
q_{2,3}^{(2)}=q_{3,2}^{(2)}=\frac{1}{\Omega}\tilde\lambda_2\tilde\lambda_3(1-N)+\frac{1}{\Omega_1}\left(\nu_{21}^p\nu_{31}^p(1-N_1)+\nu_{21}^{ap}\nu_{31}^{ap}N_1\right)+
\frac{1}{\Omega_2}\lambda_2\nu_{32}^p(1-N_2)+\\
\qquad+\frac{1}{\Omega_3}\lambda_3\nu_{13}^p\nu_{23}^p(1-N_3)\\
\vspace{1mm}
q_{3,3}^{(2)}=\frac{1}{\Omega}\tilde\lambda_3^2(1-N)+\frac{1}{\Omega_1}\left((\nu_{31}^p)^2(1-N_1)+(\nu_{31}^{ap})^2N_1\right)+
\frac{1}{\Omega_2}\left((\nu_{32}^p)^2(1-N_2)+(\nu_{32}^{ap})^2N_2\right)+\\\vspace{2mm}
\qquad+\frac{1}{\Omega_3}\lambda_3^2(1-N_3).
\end{array}%
\right.
$$

As for the $q_{k,l}^{(3)}$, these can be deduced by $q_{k,l}^{(2)}$ simply by replacing $1-N$ with $N$, $N$ with $1-N$, $1-N_j$ with $N_j$ and $N_j$ with $1-N_j$. Hence, for instance, we have
$$
q_{1,1}^{(3)}=\frac{1}{\Omega}\tilde\lambda_1^2N+\frac{1}{\Omega_1}\lambda_1^2N_1+\frac{1}{\Omega_2}\left((\nu_{12}^p)^2N_2+(\nu_{12}^{ap})^2(1-N_2)\right)+$$
$$\qquad+\frac{1}{\Omega_3}\left((\nu_{13}^p)^2N_3+(\nu_{13}^{ap})^2(1-N_3)\right),
$$
and so on.
Finally, for simple parity reasons we have $q_{k,l}^{(4)}=q_{l,k}^{(1)}$, for each $k$ and $l$.

We see that $P_j^X(t)$ in (\ref{35}) contains all that refers to the parties, while the coefficients $q_{k,l}^{(n)}$ entering in the definition of $P_j^Y(t)$ refer just to the electors, since they are all defined in terms of quantities related to $\Rc$. Hence formula (\ref{35}) clearly discriminate between the two. In principle we are now in a position to compute the various DFs for any choice of the parameters and of the initial conditions for the parties and for the electors.

\subsection{Playing with the model: the parameters $\nu^p_{kl}$ and $\nu^{ap}_{kl}$}\label{sec:play}
We  show now how the communication with the electors affects the choice of the parties regarding alliance. We shall present some different scenarios in which we see how the interaction of one party with the electors of another party  can modify the final values of the DFs. More explicitly, we describe how the various DFs are affected by the presence of non-zero $\nu^p_{kl}$ and $\nu^{ap}_{kl}$. The analysis presented extends significantly, and complete, the one just sketched in \cite{all4}, where only one particularly simple choice of the parameters was performed. Moreover, the results in \cite{all1} can also be recovered simply taking all the $\nu^p_{kl}$ and $\nu^{ap}_{kl}$ equal to zero, so that each $\Pc_j$ can only interact with $\Rc_j$ and $\Rc_{und}$, but not with $\Rc_k$, $k\neq j$.

\subsubsection{Scenario 1}\label{sectscenario1}
In the first scenario we consider the situation described by the following initial condition and parameters: $n_{1}= n_{2}= n_{3}=1$, $N_{1}=N_{3}=1, N_{2}=N=0$,
$\mu^{coop}_{12}=\mu^{coop}_{13}=\mu^{coop}_{23}=0.3$,
$\mu^{ex}_{12}=\mu^{ex}_{13}=\mu^{ex}_{23}=0.3$,
$\omega_1=\omega_2=\omega_3=\Omega_1=\Omega_2=\Omega_3=\Omega=0.1$,
$\lambda_1=0.05,\lambda_2=\lambda_3=0.01, \tilde\lambda_1=\tilde\lambda_2=\tilde\lambda_3=0.01$,  $\nu^{ap}_{12}=0.01$, $\nu^{p}_{12}\neq0$, and other parameters are set to 0. In this way, we are modeling an interaction between the parties which is {\em uniform} and stronger than the interactions between the parties and the electors. 
{ With these parameters we have carried out
three different simulations\footnote{It is clear that other choices of the parameters and of the initial conditions could be considered. In fact, we have considered many of them. What is more relevant for us is to understand the role and the importance of each parameter in the description of the system, and to check whether our interpretation is in agreement with what has been deduced for other systems, see \cite{bagbook} for instance.} with increasing values of $\nu_{12}^{p}$: 0.01, 0.1, 0.2. Because of this choice of the parameters it is clear that the strongest interactions are those between the parties, while the interactions between the parties and the electors are weaker. In fact $\mu^{ex,coop}_{ij}$'s are  larger than the $\lambda_j$'s and the $\tilde\lambda_j$. Also, the interaction between $\Pc_1$ and $\Rc_1$ is stronger than the ones between $\Pc_2$ and $\Rc_2$ or between $\Pc_3$ and $\Rc_3$. And so on. The fact that $\omega_1=\omega_2=\omega_3$  means that we are considering here parties with the same inertia, i.e. with the same a-priori attitude to form or not an alliance.
Of course, taking increasing values of $\nu_{12}^{p}$ allows us to understand better the role of $H_{mix}^p$.}


The results are presented in Fig. \ref{parnuDF1}, where it is evident
that $P_1(t)$ decreases if we increase $\nu_{12}^{p}$.  As for $P_2(t)$, which was equal to one at $t=0$, it decreases to some asymptotic value which turns out to be slightly increasing with $\nu_{12}^{p}$, see Fig. \ref{parnuDF1} (b).  In Fig.\ref{parnuDF1} (c) we see that $P_3(t)$ tends to a final asymptotic value which is { weakly} affected by $\nu_{12}^{p}$. This is because $\Pc_3$ is not directly involved in the interaction between $\Pc_1$ and $\Pc_2$ proportional to $\nu_{12}^{p}$, so that its change is just a sort of {\em second order effect}.

{It is also interesting to stress that the asymptotic values of the DFs are not dependent on their initial conditions (i.e. on what the parties want to do at $t=0$), but only on the initial attitudes of the electors. In fact, in Fig. \ref{parnuapDF2_newI} we show the time evolutions of the DFs with the (different) initial conditions $n_1=n_3,n_2=0.3$ and $\nu^{p}_{12}=0.2$, while other parameters are not changed. The selected initial conditions represent an initial picture in which none of the parties is particularly inclined to form any alliance. One can see that
the final values of the DFs are exactly those already observed in Fig. \ref{parnuDF1} for the same value of $\nu^{p}_{12}$,  $\nu^{p}_{12}=0.2$, even if the initial conditions for the parties are different. This phenomenon was already observed in \cite{all1}, in a simpler version of the system, and it is well described by the analytic forms of the DFs given there (see also (\ref{33b}) for a similar formula with a peculiar choice of the parameters). This result shows the relevance of the interactions with the electors, which is what really drives the parties toward their final decision, regardless of the initial interest of the parties to ally. Then the parties are {\em perfect}, i.e. they really listen to their electors.

\begin{figure}[!]
			\begin{center}
			\subfigure[$P_{1}(t)$]{\hspace*{-0.2cm}\includegraphics[width=7cm]{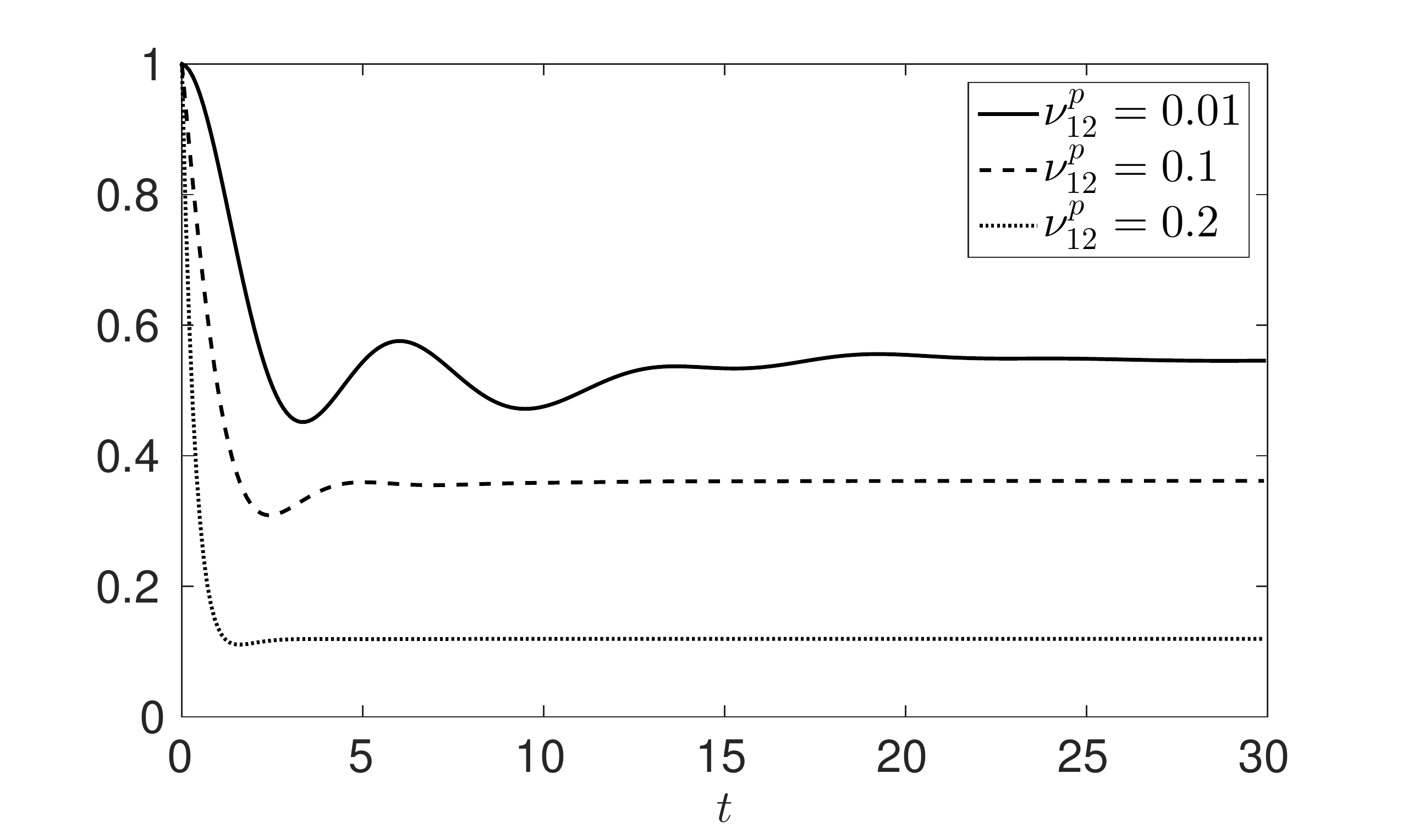}}
	      	\subfigure[$P_{2}(t)$]{\hspace*{-0.1cm}\includegraphics[width=7cm]{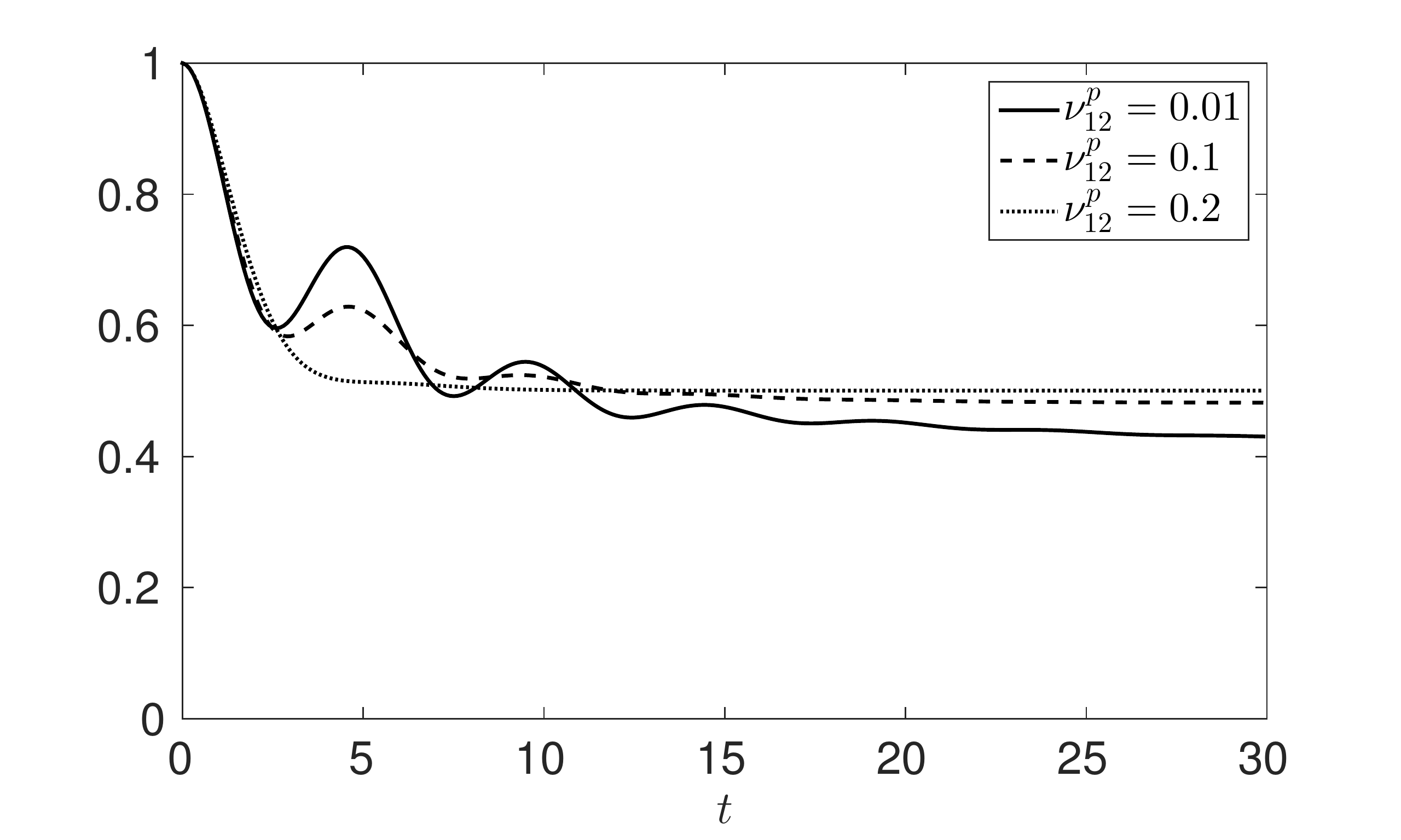}}	
			\subfigure[$P_{3}(t)$]{\hspace*{-0.1cm}\includegraphics[width=7cm]{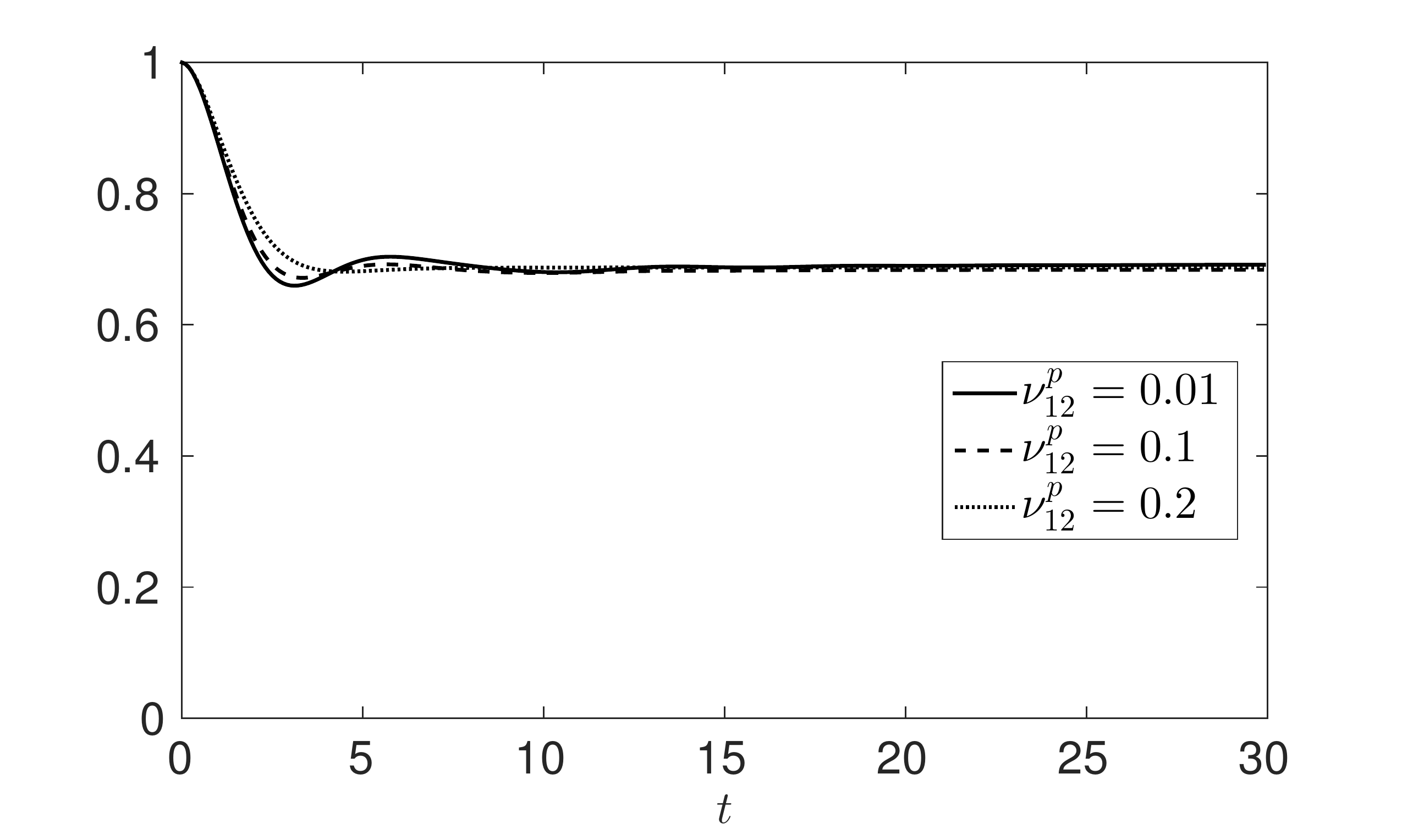}}	
			\end{center}
\vspace*{-0.5cm}\caption{ Time evolution of the DFs for various values of $\nu^p_{12}$. The initial conditions  are $n_{1}= n_{2}= n_{3}=1$, $N_{1}=N_{3}=1, N_{2}=N=0$. The choice of the other parameters is the following:
$\mu^{coop}_{12}=\mu^{coop}_{13}=\mu^{coop}_{23}=0.3$,
$\mu^{ex}_{12}=\mu^{ex}_{13}=\mu^{ex}_{23}=0.3$,
$\omega_1=\omega_2=\omega_3=\Omega_1=\Omega_2=\Omega_3=\Omega=0.1$,
$\lambda_1=0.05,\lambda_2=\lambda_3=0.01, \tilde\lambda_1=\tilde\lambda_2=\tilde\lambda_3=0.01$, $\nu_{12}^{ap}=0.01$. The other parameters are set to 0.}
\label{parnuDF1}
\end{figure}

\begin{figure}[!]
			\begin{center}
				\includegraphics[width=8cm]{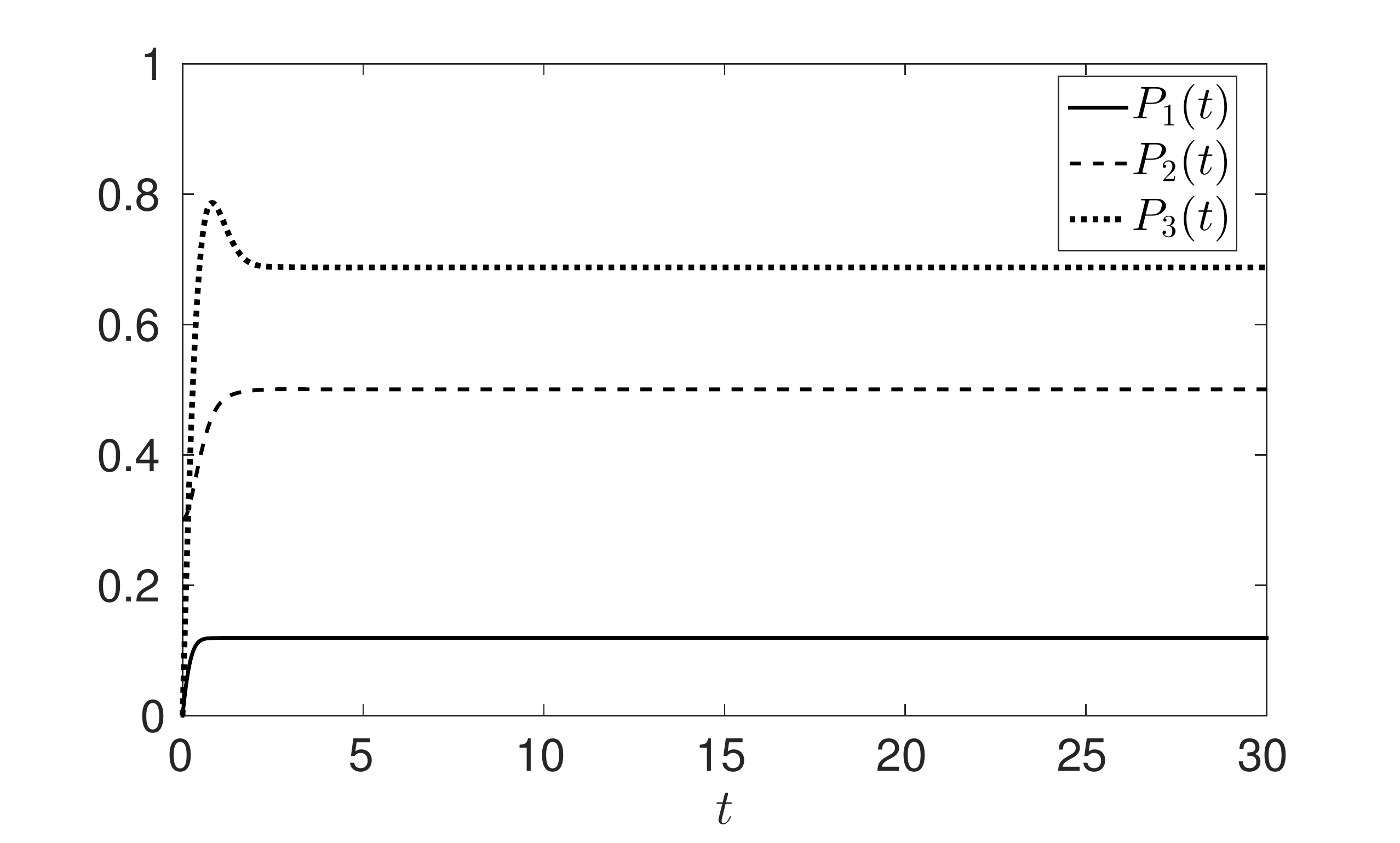}
			\end{center}
\vspace*{-0.5cm}\caption{Time evolution of the DFs for the initial conditions $n_1=0,n_2=0.3,n_3=0$ and other parameters as in Fig. \ref{parnuDF1} with $\nu_{12}^{p}=0.2$.
}
\label{parnuapDF2_newI}
\end{figure}

\subsubsection{Scenario 2}
In the second scenario, we consider the effects due to the variation of the parameter  $\nu_{12}^{ap}$.  We consider the same initial conditions as in the first part of the previous scenario, we fix $\nu_{12}^{p}=0.01$ and we choose the following increasing values for $\nu_{12}^{ap}$: 0.01 (already analyzed in the first scenario), 0.15 and 0.35. The results are presented in Fig. \ref{parnuapDF2}. We observe that $P_1(t)$ tends asymptotically to a value which is closer and closer to 1
as $\nu_{12}^{ap}$ increases. Analogously to what we have seen in Section \ref{sectscenario1}, we observe that the asymptotic value  $P_2(\infty)$ slightly increases for higher
$\nu_{12}^{ap}$, whereas $P_3(t)$ tends to an asymptotic value which is almost independent of  $\nu_{12}^{p}$, for the same reason as before.
 In Fig. \ref{parnuapDF3_newI} we plot the  time evolution of the DFs for initial conditions $n_1=n_3=0$, $n_2=0.3$, different from those used in Fig. \ref{parnuapDF2}, and for $\nu^{ap}_{12}=0.35$. The other parameters are the same as in Fig. \ref{parnuapDF2}. We see that
the asymptotic values  are exactly those already deduced in Fig. \ref{parnuapDF2} for the same value of  $\nu^{ap}_{12}=0.35$. Again, the initial ideas of the parties are irrelevant for their final decision.

\begin{figure}[!]
			\begin{center}
				 \subfigure[$P_{1}(t)$]{\hspace*{-0.1cm}\includegraphics[width=7cm]{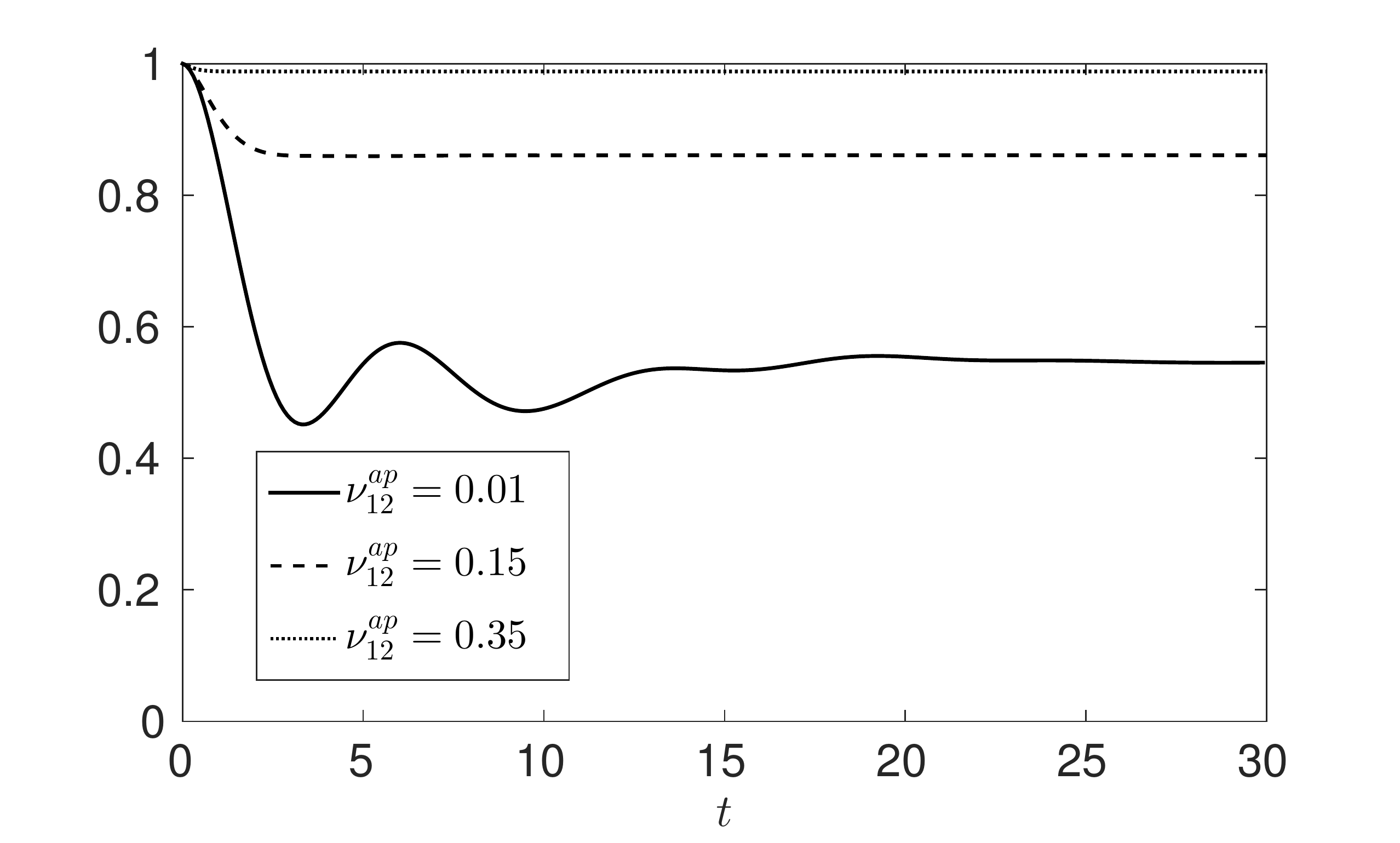}}
				 \subfigure[$P_{2}(t)$]{\hspace*{-0.1cm}\includegraphics[width=7cm]{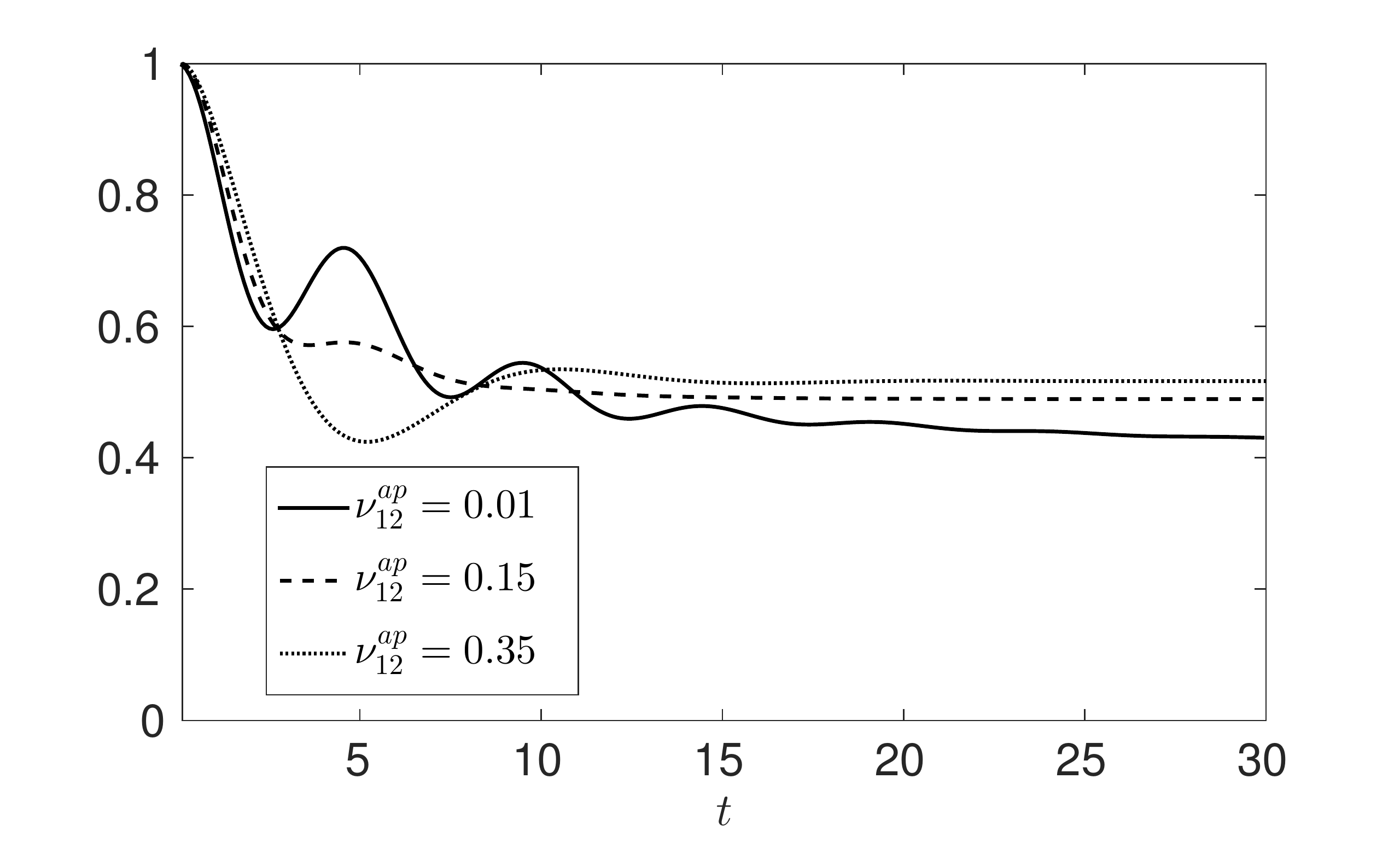}}	
				 \subfigure[$P_{3}(t)$]{\hspace*{-0.1cm}\includegraphics[width=7cm]{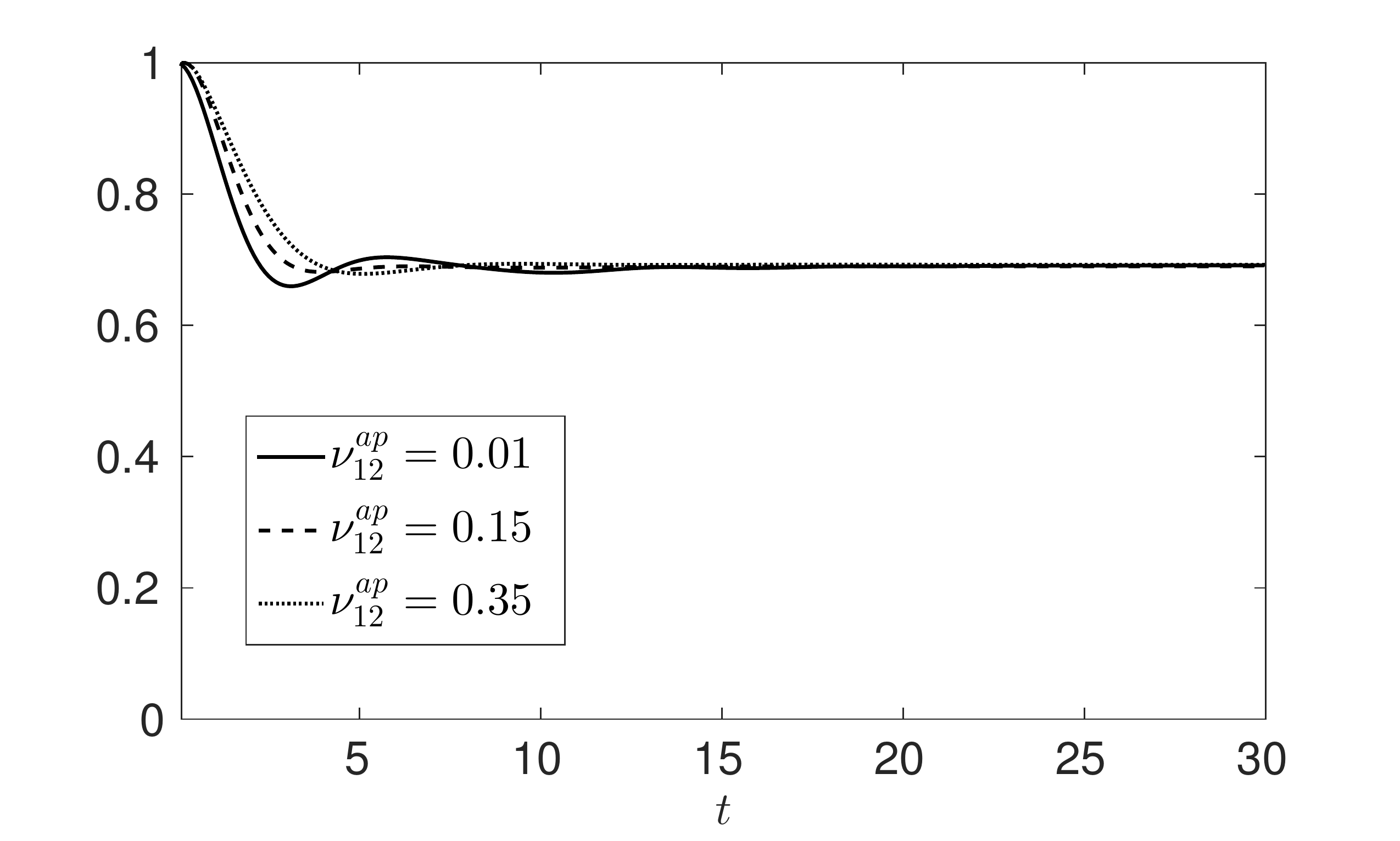}}	
			\end{center}
\vspace*{-0.5cm}\caption{Time evolution of the DFs for various values of $\nu^{ap}_{12}$, and for $\nu_{12}^{p}=0.01$. The initial conditions and the other parameters are as in Fig. \ref{parnuDF1}. }
\label{parnuapDF2}
\end{figure}

\begin{figure}[!]
			\begin{center}
				\includegraphics[width=8cm]{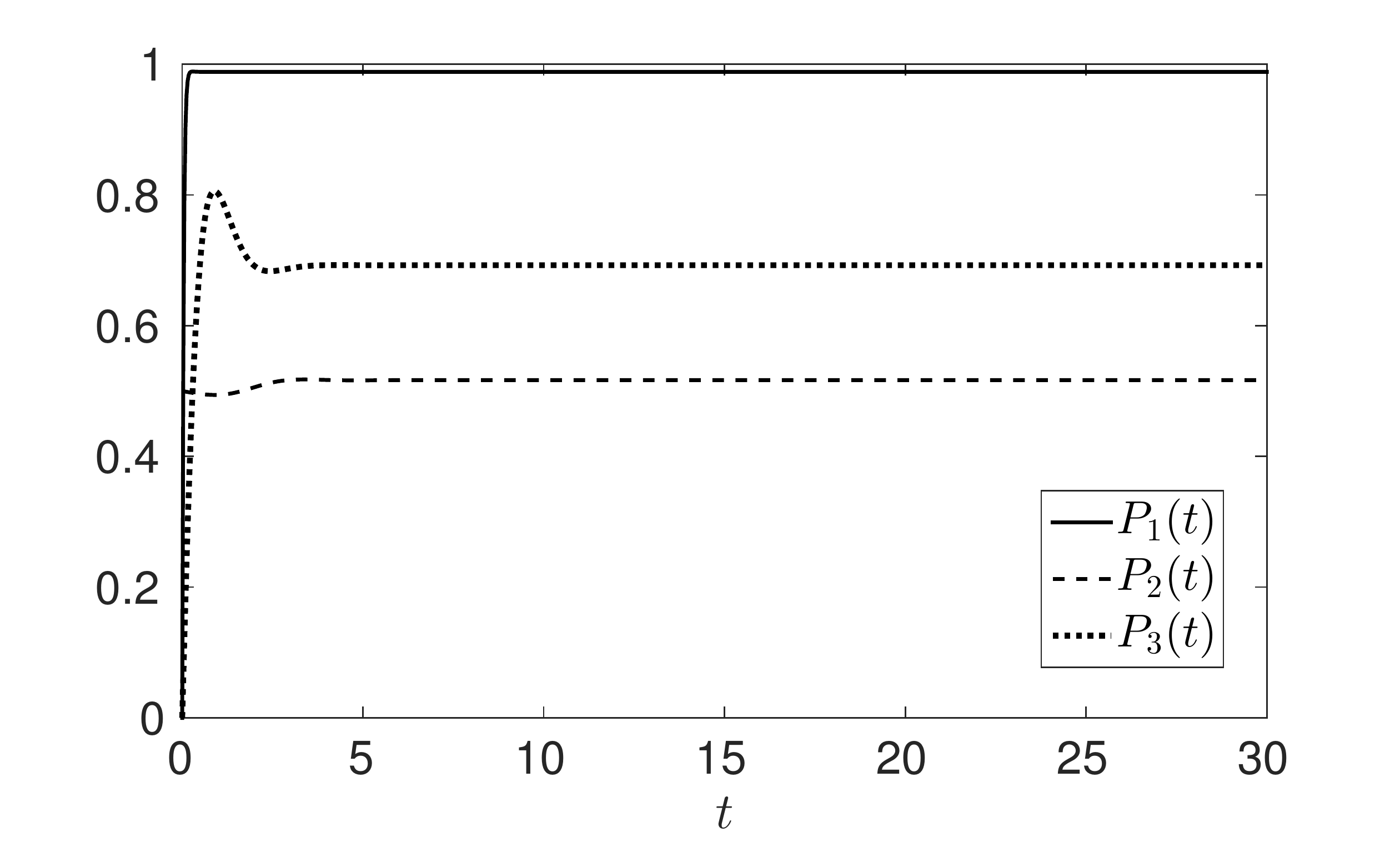}
			\end{center}
\vspace*{-0.5cm}\caption{Time evolution of the DFs for the initial conditions $n_1=0,n_2=0.3,n_3=0$ and other parameters as in Fig. \ref{parnuDF1}, with $\nu_{12}^{ap}=0.35$.}
\label{parnuapDF3_newI}
\end{figure}

\subsubsection{Scenario 3}

 In this last scenario we analyze the effect of the simultaneous presence of two non-zero parameters in the Hamiltonian measuring interactions of different  nature. In particular we modify the parameters $\lambda_1$, which tunes the strength of the interaction between $\Pc_1$ and $\Rc_1$, and  $\nu^p_{12}$, which measures the interaction between $\Pc_1$ and $\Rc_2$. If we take both the parameters of the same order of magnitude, we expect that $P_1(t)$ will be influenced by both $N_1$ and $N_2$. This could mimic the presence of some kind of fragmentation within the members of the party $\Pc_1$, because a part of its members is more inclined to follow what suggested by their own electors, while other members could be more inclined to follow the suggestion of the electors in $\Rc_2$. Obviously, this becomes really interesting when $N_1$ and $N_2$ are sufficiently different, since otherwise this difference is not particularly evident. When this happens, the obvious question would be: it is more convenient to listen to its own electors, or to follow what other electors are suggesting?

To be concrete, we consider  the following choice of initial conditions and parameters:
$n_{1}= n_{3}=0, n_{2}=0.3$, $N_{1}=N_{3}=1, N_{2}=N=0$,
$\mu^{coop}_{12}=\mu^{coop}_{13}=\mu^{coop}_{23}=0.3$,
$\mu^{ex}_{12}=\mu^{ex}_{13}=\mu^{ex}_{23}=0.1$,
$\omega_1=\omega_2=\omega_3=0.1,\Omega_1=\Omega_2=\Omega_3=0.4,, \Omega=0.05$,
$\lambda_2=0.01,\lambda_3=0.3, \tilde\lambda_1=\tilde\lambda_2=\tilde\lambda_3=0.01$, $\nu_{12}^{p}=0.2$, $\lambda_1$ varies between 0.2 and 0.8, and the other parameters are set to 0.

{ This choice of the parameters describes a strong interaction between $\Pc_1$ and both $\Rc_1$ and $\Rc_2$ (due respectively to the high values of  to $\lambda_1$ and $\nu_{12}^{p}$), while $\Pc_2$ and $\Pc_3$ are only interacting  with their own electors, but not with the electors of the other parties. This choice is meant to put some asymmetry in the system. Due to the contrasting initial conditions $N_1=1$, $N_2=0$, we expect that $\Pc_1$ experiences some kind of indecision when $\lambda_1=\nu_{12}^{p}$, while, for increasing  $\lambda_1$, the final decision of $\Pc_1$ should be more and more affected by the feeling of $\Rc_1$.
Due to the cooperative and competitive interactions  ($\mu^{ex,coop}_{ij}\neq0$), $\Pc_2$ and $\Pc_3$ interact with $\Pc_1$, and hence we expect that also $P_2,P_3$ should be somewhat influenced by the variation of $\lambda_1$.
In particular, since the $\mu^{coop}_{ij}$'s are greater than the $\mu^{ex}_{ij}$'s, the parties have strong cooperation, hence it is expected that when increasing $\lambda_1$, the DFs $P_2(t)$ and $P_3(t)$ should stay closer to $N_1$.
}


The results are shown in Fig. \ref{parlambdaDF} for various values of $\lambda_1$. When $\lambda_1=\nu^p_{12}=0.2$,  there is balance between the two effects and $P_1(t)$ tends toward a value which is almost $0.513$, very close the mean value of $N_1$ and $N_2$. If we increase $\lambda_1$ with respect to $\nu^p_{12}$ this balance is lost, and $P_1(t)$ increases, meaning that  $\Pc_1$ is  more inclined to follow what their electors $\Rc_1$ are suggesting. As expected, a weak change with $\lambda_1$ is also observed in the plots of $P_2(t)$ and $P_3(t)$.

\begin{figure}[!]
			\begin{center}
				 \subfigure[$P_{1}(t)$]{\hspace*{-0.1cm}\includegraphics[width=7cm]{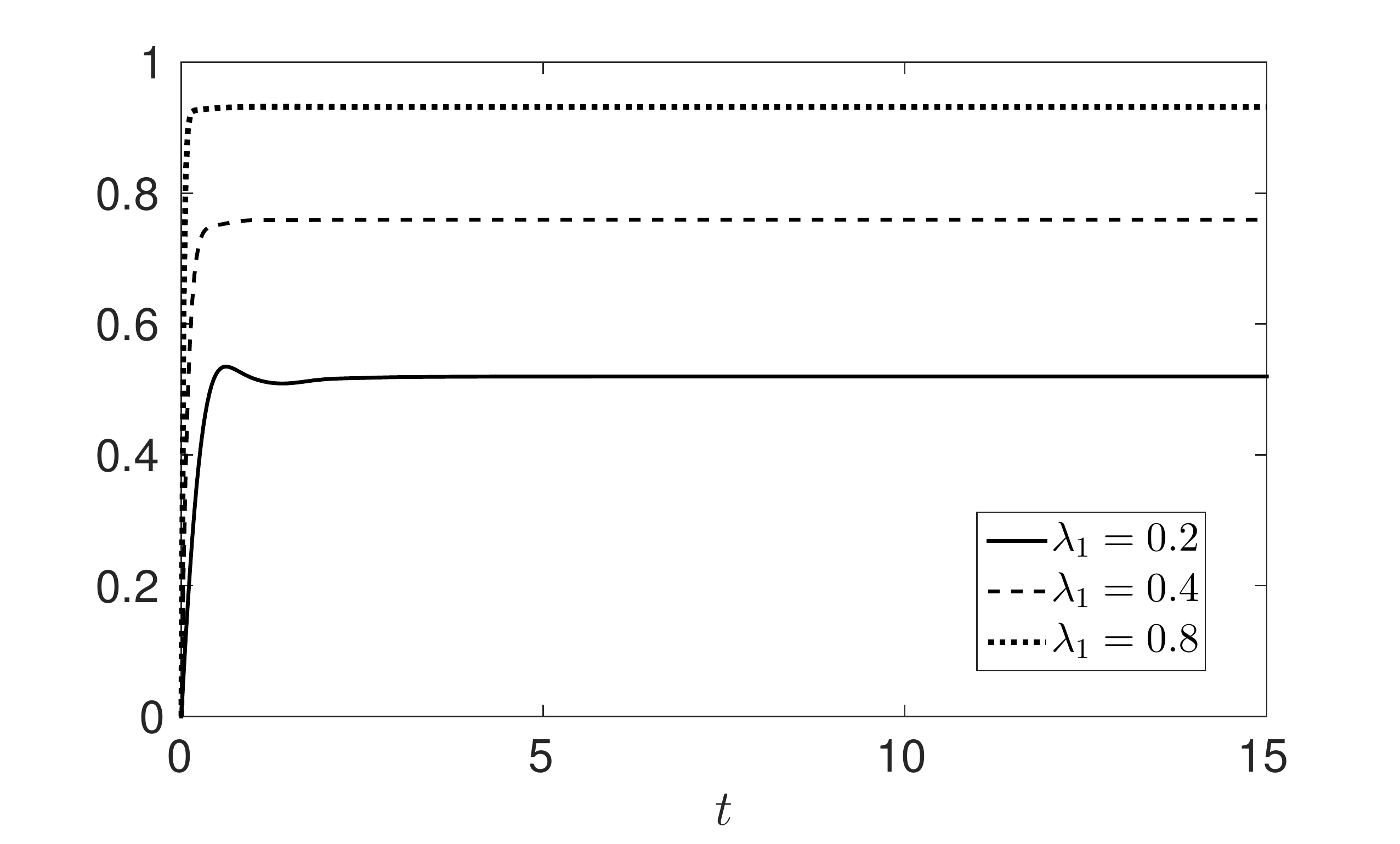}}
				 \subfigure[$P_{2}(t)$]{\hspace*{-0.1cm}\includegraphics[width=7cm]{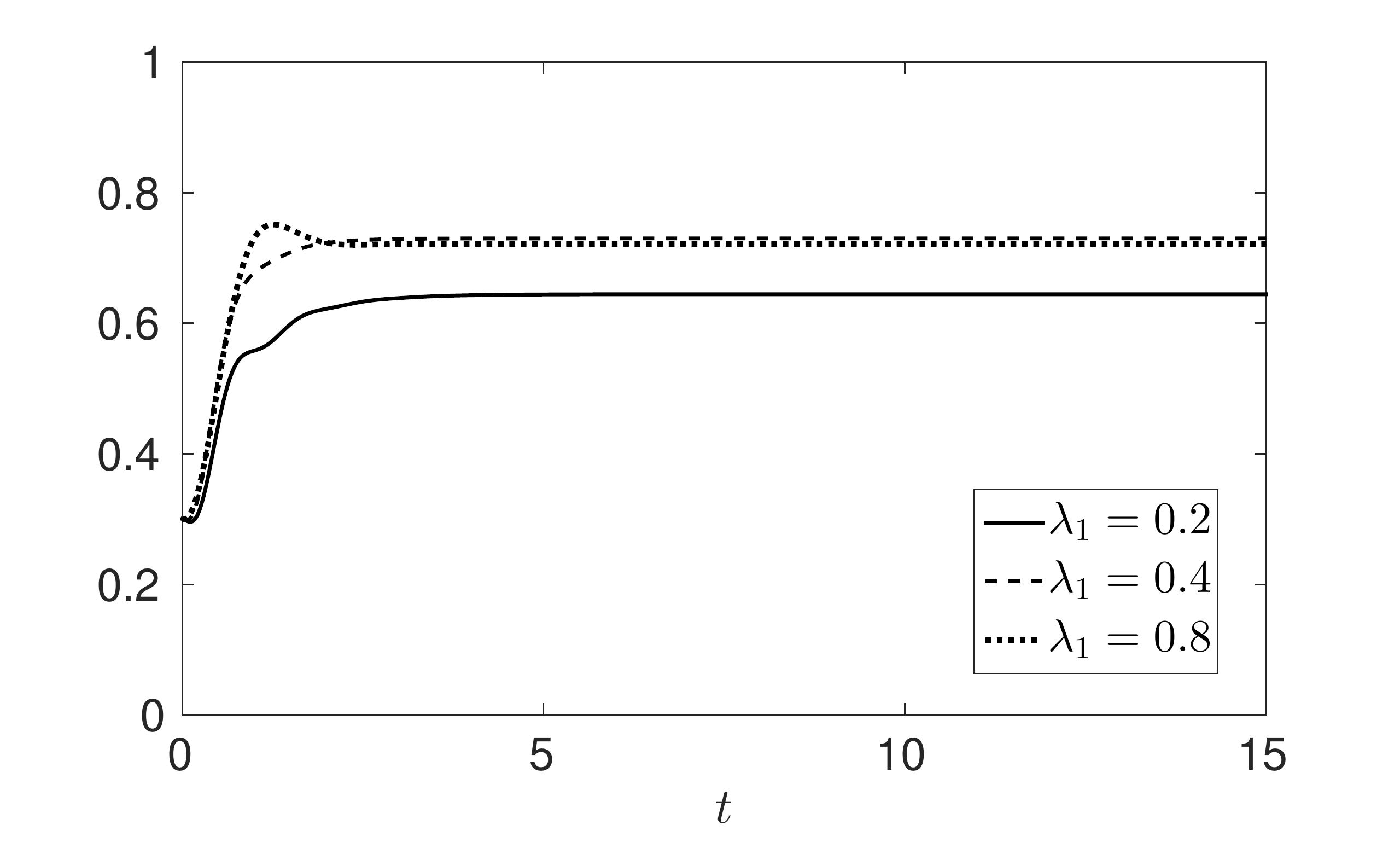}}	
				 \subfigure[$P_{3}(t)$]{\hspace*{-0.1cm}\includegraphics[width=7cm]{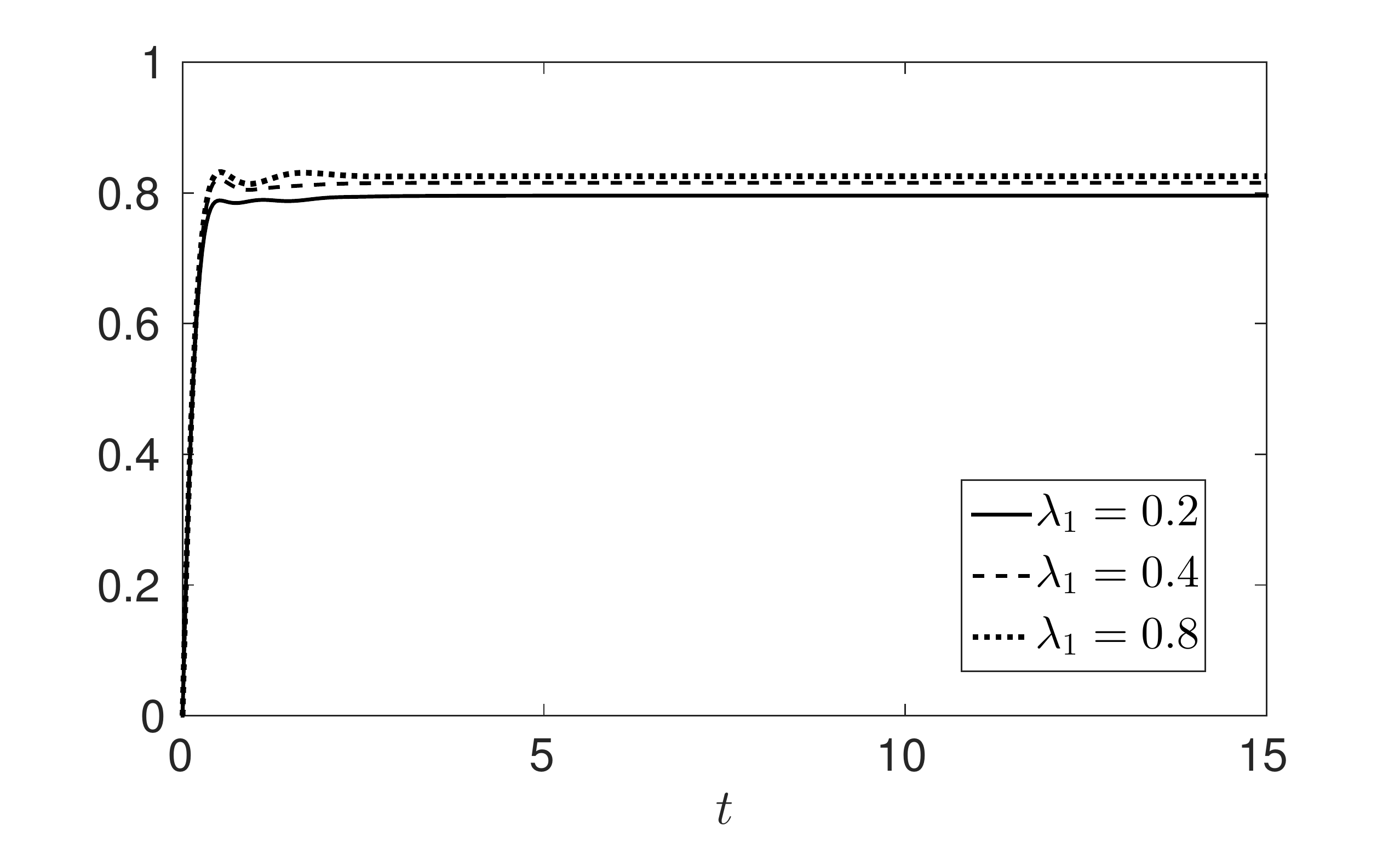}}	
			\end{center}
\vspace*{-0.5cm}\caption{Time evolution of the DFs for various values of $\lambda_{1}$. Initial conditions are  $n_1=0,n_2=0.3,n_3=0$, $N_{1}=N_{3}=1, N_{2}=N=0$ and the parameters are $\mu^{coop}_{12}=\mu^{coop}_{13}=\mu^{coop}_{23}=0.3$,
$\mu^{ex}_{12}=\mu^{ex}_{13}=\mu^{ex}_{23}=0.1$,
$\omega_1=\omega_2=\omega_3=0.1,\Omega_1=\Omega_2=\Omega_3=0.4,\Omega=0.05$,
$\lambda_2=0.01,\lambda_3=0.3, \tilde\lambda_1=\tilde\lambda_2=\tilde\lambda_3=0.01$, $\nu_{12}^{p}=0.2$.}
\label{parlambdaDF}
\end{figure}

\vspace{3mm}

Other simulations can be found in\cite{all1} and \cite{all4}. In particular, it might be worth to remind that, when the parties only interact among themselves, the DFs have been found in \cite{all1} to oscillate reaching no asymptotic value: it is only the interaction with the electors which produces a {\em real} decision!

\section{Adding a {\em rule} $\rho$ to the model}\label{sectrule}

We introduce now an important variation  of the  model considered so far. This variation is based on the introduction of some \emph{rules} that somehow correct the Heisenberg dynamics. The application of  rules in the framework of physical systems described by quantum tools has been recently proposed in \cite{BDGO}, where the notion of  $(H,\rho)$-induced dynamics has been introduced and applied to the description of the time evolution of a population  in a bounded 2D region and in the so-called quantum game of life. Some basic facts of this induced dynamics can be found in the Appendix.

Here we suppose that at each time step $\tau$, some kind of information reaches the electors, which can then modify their opinions on whether their party should form an alliance or not. In other words,  we suppose that the values $N_1$, $N_2$ and $N_3$ in (\ref{211}), and possibly $N$ in (\ref{211bis}), can be modified at each time step $\tau$ because of this information. Then, with these new values, the time evolution starts again for a second time interval of length $\tau$, when the rule is applied again. And so on. This is, in synthesis, our rule. We will make it more precise and explicit later in this section. In particular,
we set the initial conditions $P_{j,0}(0)=n_{j,0}$ for the DFs, and the initial condition  $(N_{j,[0]},N_{[0]})$ for the reservoirs\footnote{In this general introduction to the rule, we will consider the possibility that both $\Rc_j$ and $\Rc_{und}$ are modified by the rule. However, in the concrete situation described by (\ref{rules_eq}), $\Rc_{und}$ is not changed at all by $\rho$.}, and we fix the {\em  time step} $\tau$ such that in the time interval $[0,\tau]$ the time evolution of each DF $P_{j,0}(t)$ is deduced, given the Hamiltonian \eqref{31}, through \eqref{35}.
Notice that we are now using two subscript indices $(j,k)$ since the first is related to the party ($j=1,2,3$) while the second labels  the number of the iteration (the first iteration is set to 0).
At $t=\tau$ the values $P_{j,0}(\tau)$ are taken as input for a set of rule $\rho$ (specified later)  which returns as output the new values $(N_{j,[1]},N_{[1]})$, in general different from $(N_{j,[0]},N_{[0]})$, for the reservoirs $(\Rc_j,\Rc_{und})$; this is because, at time $\tau$,
 the electors may react in different ways depending on whether they are satisfied or not with what the parties are doing.
Then we start a new iteration with the initial conditions
$P_{j,1}(0)=P_{j,0}(\tau)$ for the DFs and with $(N_{j,[1]},N_{[1]})$ for the reservoirs, and the evolution of the DFs is again ruled by the Hamiltonian \eqref{31} in a second time interval of length $\tau$. Then $\rho$ is applied a second time, and $(N_{j,[2]},N_{[2]})$ are deduced.
This process can be iterated   $\mathcal{N}$ times so that, at the end, we have obtained a sequence of DFs $P_{j,k}(t)$,
 $j=1,2,3$, $k=0\ldots \mathcal{N}-1, t\in[0,\tau]$. Hence the global DFs, in the time interval $[0,\mathcal{N}\tau[$, are reconstructed through
\be
P_j(t)=\sum\limits_{k=0}^{\mathcal{N}-1} \chi_k(t) P_{j,k}(t-k\tau),\quad j=1,2,3,\label{wDFs}
\en
where $\chi_k(t)=1$ if {$t\in[k\tau,(k+1)\tau[$}, and $\chi_k(t)=0$ otherwise. Notice that the DFs  are continuous (but not differentiable, in general) in time.

\subsection{Making the rule}
Of course, the key ingredient for the whole procedure described above is the concrete definition of the rule $\rho$. As specified before, at each iteration,
the rule works by taking as input the values $(P_{j,k}(\tau),N_{j,[k]},N_{[k]})$ and giving as output $(N_{j,[k+1]},N_{[k+1]})$. Needless to say, this procedure is not unique. We describe now the one we have used in this paper, which is based on the following general ideas:
\begin{enumerate}
\item The values $(N_{j,[k]},N_{[k]})$ are considered as a strength measuring how strongly the electors suggest the parties to ally or not: if, for instance, $N_{1,[k]}\approx1$, then  the electors in $\mathcal{R}_1$  are strongly suggesting  $\mathcal{P}_1$ to form some alliance, while if, for instance, $N_{1,[k]}\approx0.6$, then
$\mathcal{R}_1$  is still suggesting $\mathcal{P}_1$ to ally, but in a {\em milder} way.
\item If a party {follows} the suggestions of its electors (that is $P_{j,k}\approx1$ when $N_{j,[k]}\approx1$ or $P_{j,k}\approx0$ when $N_{j,[k]}\approx0$), then we do not expect the electors to change much their behaviors after the rule is applied. In other words,
$(N_{j,[k+1]},N_{[k+1]})$ should not be particularly different from $(N_{j,[k]},N_{[k]})$.

\item On the other hand, when $\Pc_j$ is not doing what $\Rc_j$ is suggesting to do, $N_{j,[k+1]}$ becomes very close to one or zero (depending on whether $\Rc_j$ wants $\Pc_j$ to form or not an alliance), independently of the value of $N_{j,[k]}$. Hence, in a single iteration, $|N_{j,[k+1]}-N_{j,[k]}|$ could turn out to be close to one, which is the maximum value allowed in our scheme.

\item The speed of change of the various DFs should be somehow related to the strength of the input coming from the electors, i.e. to the values of the $(N_{j,[k]},N_{[k]})$.

\item The undecided electors are not affected by what the parties are doing: they simply don't care about the parties! Then  $N_{[k+1]}=N_{[k]}$, for all $k$.
\end{enumerate}

{As already stated, other possible {\em general ideas} could be assumed, possibly as reasonable as those listed here. For instance, we could also consider some effect of the rule on the undecided electors. But, at least in this paper, we will work with this list.}

Following these simple but realistic requirements, we define the  rule $\rho$ to determine $N_{j,[k+1]}$ as follows:

\be
N_{j,[k+1]}=\rho(N_{j,[k]},\delta_j,P_{j,k}(\tau))=\left\{
\begin{array}{ll}
  N_{j,[k]} +(1-P_{j,k}(\tau))^{s} (1 - N_{j,[k]})\frac{(1 - \delta_j)}{(1 + |\delta_j|)}, \quad \textrm{if } N_{j,[k]}\geq0.5\\
 N_{j,[k]} - (P_{j,k}(\tau))^sN_{j,[k]}\frac{(1 + \delta_j)}{(1 + |\delta_j|)}, \quad\qquad\qquad \textrm{if } N_{j,[k]}<0.5,
\end{array}
\right.
\label{rules_eq}
\en
where $\delta_j=P_{j,k}(\tau)-P_{j,k}(0)$, and $s$ is a non negative parameter. First we observe that, using (\ref{rules_eq}), the new values $N_{j,[k+1]}$ stay always in the range $[0,1]$. Now, in order to better understand (\ref{rules_eq}), and to show the relation between (\ref{rules_eq}) and the five general ideas listed above, we start noticing that the rule does not change the value of $N_{[k]}$. This is point 5 above.

{Now, let us consider first what happens if $N_{j,[k]}\geq0.5$, i.e., if the electors in $\Rc_j$ are suggesting $\Pc_j$ to form some alliance. Of course, the closer $N_{j,[k]}$ is to one, the stronger this suggestion. If the party $\Pc_j$ is not following what suggested by its electors, then $\delta_j<0$. This is because, in this case, the electors in $\Rc_j$ would like $\Pc_j$ to ally (remember that $N_{j,[k]}\geq0.5$), while the DF of $\Pc_j$ is decreasing. Then $(1 - \delta_j)/(1 + |\delta_j|)=1$ and
$N_{j,[k+1]}=N_{j,[k]} +(1-P_{j,k}(\tau))^{s} (1 - N_{j,[k]})$. Hence $N_{j,[k+1]}$ is surely larger than $N_{j,[k]}$, and the term $(1-P_{j,k}(\tau))^{s}$ tunes how rapidly $N_{j,[k+1]}$ is increasing. The fact that $N_{j,[k+1]}$ cannot be larger than one is due to the presence of the factor $(1 - N_{j,[k]})$, which ensures that, at most, $N_{j,[k+1]}$ reaches the value one. {This fulfills the requirements of point 3 above. Otherwise, if $\Pc_j$ is following what its electors are suggesting, then $\delta_j>0$.
 In this case $(1 - \delta_j)/(1 + |\delta_j|)\in ]0,1[$, so that
 $N_{j,[k+1]}$ still increases with respect to $N_{j,[k]}$, but with a rate which also depends on the explicit value of $\delta_j$. Similarly one could explain the rule in the case $N_{j,[k]}<0.5$, where the only (essential) difference is that now the electors are suggesting $\Pc_j$  not to ally at all.



 Although the above rule seems sufficiently realistic, it is clear that other possibilities could also be considered. For instance, another reasonable rule $\tilde\rho$, which will not be considered here, is the one which modifies, at each step $\tau$, the parameters of the hamiltonian $h$ in (\ref{31}) according to some external factors, like, for instance, some opinion polls.

\subsection{Concrete applications of the rule}
In this section we consider some applications of the rule in (\ref{rules_eq}). In all cases discussed here the parameter $s$  is fixed to be $\frac{1}{4}$. The time $\tau$ is taken to be $25$, and we consider just four iterations.

\subsubsection{}\label{sec:caserule1} The first application is related to a simple case, in which the various parties interact only with their electors, but not with the other parties and with the other electors. This case, in absence of any rule, was considered in \cite{all1}, and corresponds to the following choice of the parameters $\mu^{ex}_{k,l}=\mu^{coop}_{k,l}=\tilde\lambda_k=\nu^p_{k,l}=\nu^{ap}_{k,l}=0$, for all $k$ and $l$.
It is easy to check that in this case the matrix $U$ is diagonal, and the DFs, at each iteration, have the following simple expression:

\be\label{33b}
P_{j,k}(t)=P_{j,k}(0)e^{-2\pi t \lambda_j^2/\Omega_j}+N_{j,[k]}\left(1- e^{-2\pi t \lambda_j^2/\Omega_j} \right).
\en
Of course, this equation implies  that $P_{j,k}(t)$ would tend asymptotically (i.e., for $t$ diverging\footnote{But recall that here $t\leq\tau$, and $\tau=25$!}) to $N_{j,[k]}$ with a convergence speed which depends on the value of $\lambda_j^2/\Omega_j$.

The DFs deduced after the four iterations, \eqref{wDFs}, are shown in Fig. \ref{rulesFIG1}, where we have taken the initial conditions as follows:  $n_{1,0}=0.2, n_{2,0}=0.8, n_{3,0}=0.5$, $N_{1,[0]}=0.1, N_{2,[0]}=0.6, N_{3,[0]}=1, N_{[0]}=1$ and we have considered the following choice of the other parameters:
$\omega_1=\omega_2=0.1,\omega_3= 0.2, \Omega_1=\Omega_2=\Omega_3=\Omega=0.1, \lambda_1=0.08,\lambda_2=\lambda_3=0.02$. This means that, at $t=0$,
$\mathcal{P}_1$  is
not inclined to ally ($n_{1,0}=0.2$), and that its electors, those in $\Rc_1$, do not want $\Pc_1$ to form any alliance ($N_{1,[0]}=0.1$).
Actually, in absence of any rule, formula (\ref{33b}) predicts that  $P_{1,0}(t)$ tends to 0.1 when $t\rightarrow\infty$. However, applying the rule \eqref{rules_eq} at $t=\tau$, we find a new value $N_{1,[1]}$, which is 0.0181, quite below the previous value $N_{1,[0]}=0.1$. This means that the electors $\Rc_1$ are now suggesting $\Pc_1$  not to ally in a even stronger way than before. At the end of the second iteration, at $t=50$, $P_{1}(50)=P_{1,1}(25)=0.018$, and by applying the rule we obtain
$N_{1,[2]}=0.00275$, still quite below than the previous value $N_{1,[1]}$.
By continuing this procedure, at the end of the fourth iteration, the asymptotic value of the DF for $\mathcal{P}_1$ is $P_1(100)=0.0000835$.
It is natural to imagine that, continuing to apply the rule for more and more iterations, we would indeed find $P_{1}(t)\rightarrow0$ as $t\rightarrow\infty$: we see that  $\Pc_1$ is  doing what $\Rc_1$ was suggesting!

Concerning the  party $\mathcal{P}_2$, the initial condition  $N_{2,[0]}=0.6$ means that the electors in $\Rc_2$ are (gently) suggesting $\Pc_2$ to ally with some other party. However, as deduced by the formula in (\ref{33b}), the DF $P_{2,0}(t)$  tends to $N_{2,[0]}=0.6$ for $t$ large, which is a value below its initial value, $P_{2,0}(0)=0.8$.
In fact, already for $t=25$, we obtain $P_{2,0}(25)=0.7$, which is smaller than the initial value 0.8.
This means that the party is partially going against what their electors $\Rc_2$ are suggesting: $\Pc_2$ is decreasing its original tendency to form an alliance, even if $\Rc_2$ is (gently) suggesting to do that. However, applying the rule at $t=\tau$,
we get a new value for $N_{2,[1]}$ which is 1, much larger than $N_{2,[0]}$. This shows that the electors are reacting in a \textit{bad} way against their party, because $\Pc_2$ is not listening to them.  In the successive  iterations the new values of $N_{2,[k]}$ remains equal to 1, and the function $P_{2}(t)$ continues to increase. This is clearly shown in Fig. \ref{rulesFIG1}.

Finally, regarding the party $\mathcal{P}_3$, there is not  much to say: in fact their electors $\Rc_3$ are suggesting to ally because $N_{3,[0]}=1$. Hence the DF $P_3(t)$ can only increase in time toward the asymptotic value $1$ and this is, in fact, what is observed in Fig. \ref{rulesFIG1}.

\begin{figure}
			\begin{center}
		\hspace*{-0.2cm}\includegraphics[width=8cm]{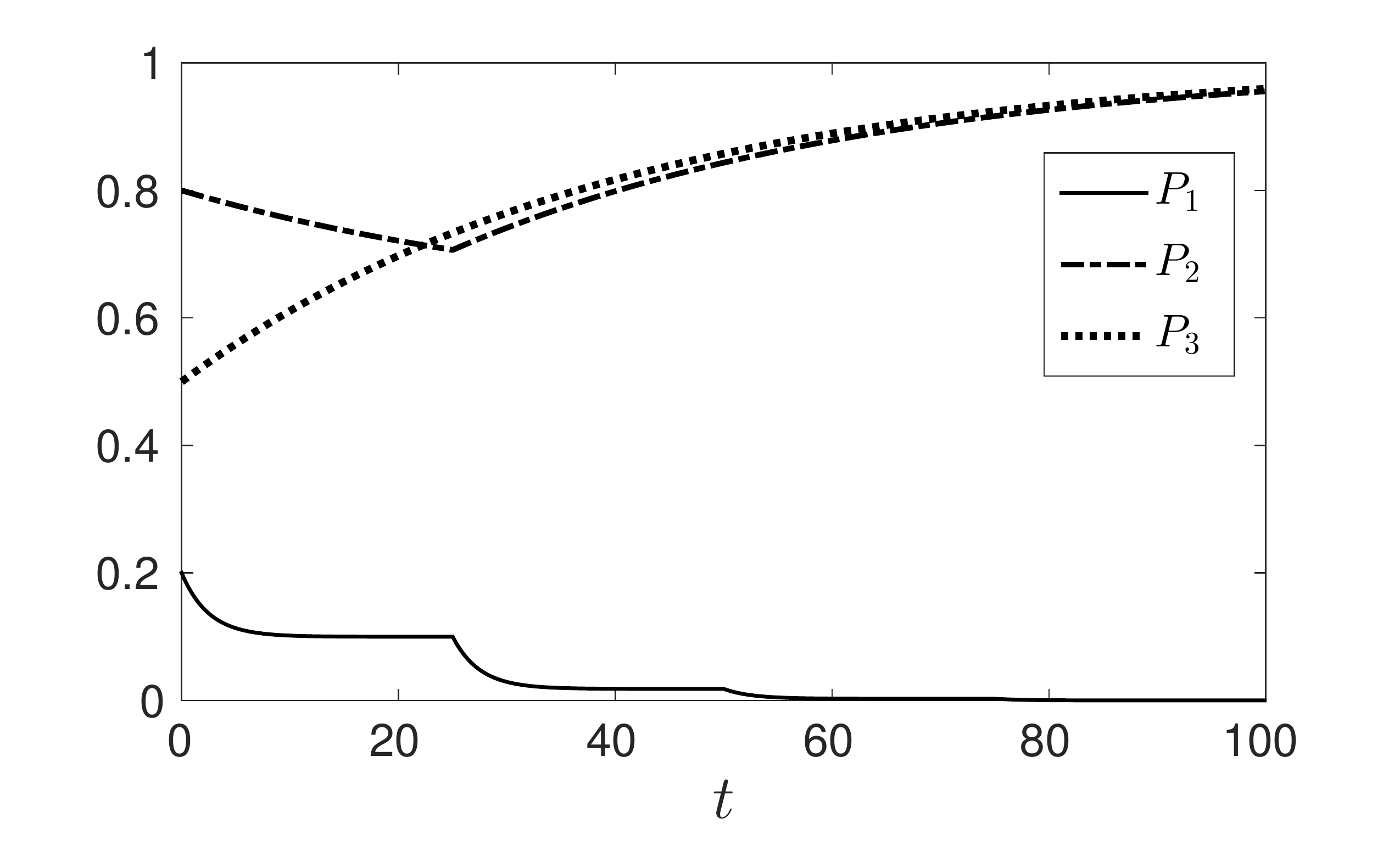}
			\end{center}
\vspace*{-0.5cm}\caption{ Initial conditions are: $n_{1,0}=0.2, n_{2,0}=0.8, n_{3,0}=0.5$, $N_{1,[0]}=0.1, N_{2,[0]}=0.6, N_{3,[0]}=1, N_{[0]}=1$. The choice of the parameters is the following:
$\omega_1=\omega_2=0.1,\omega_3= 0.2, \Omega_1=\Omega_2=\Omega_3=\Omega=0.1, \lambda_1=0.08,\lambda_2=\lambda_3=0.02$. The other parameters are set to 0.
}
\label{rulesFIG1}
\end{figure}

\subsubsection{}\label{sec:caserule2}

In this second application we add the effect of the communication between the parties and the other parties' electors, so that the parameters $\lambda_j,\tilde\lambda_j,\nu^p_{kl},\nu^{ap}_{kl}$ can be different from 0.
In particular we put $\lambda_1=0.08,\lambda_2=\lambda_3=0.02, \tilde\lambda_1=0.01$ and $\nu^p_{12}=\nu^{ap}_{12}=\nu^{ap}_{13}=0.001$. The other parameters and the initial conditions are fixed as in Subsection \ref{sec:caserule1}.

It is evident that with the choice of these parameters the  party $\Pc_1$ has the strongest interactions with their own electors, hence we expect that
$\Pc_1$ should decide following essentially what its electors in $\Rc_1$ are suggesting, that is no alliance at all. However,  the choice $\nu^p_{12},\nu^{ap}_{12},\nu^p_{13}\neq0$ implies that $\Pc_1$ is also (but less!) influenced by the electors in $\Rc_2$ and $\Rc_3$.
The results are shown in Fig. \ref{rulesFIG2}. Although the various  $N_{1,[k]}$ tend to zero
($N_{1,[1]}=0.0053,N_{1,[2]}=0.0008,N_{1,[3]}=6\cdot 10^{-6}$), the DF of $\Pc_1$ seems to stabilize around the value 0.085. This expresses the fact that,
regardless the attitude of the electors $\Rc_1$, the party $\Pc_1$ can slightly modify its attitude taking into account also what $\Rc_2$ and $\Rc_3$ are suggesting. In other words, in this case the decision is not as sharp as in absence of the (mixed) interactions between $\Pc_1$ and $\Rc_2$ and $\Rc_3$.

Conversely,  the electors $\Rc_2$ and $\Rc_3$  suggest  $\Pc_2$ and $\Pc_3$ to ally (even if with different strengths), since we have $N_{2,[0]}=0.6$ and $N_{3,[0]}=1$. Applying the rule, at the various iterations we obtain
$N_{2,[k]}=N_{3,[k]}=1$, $k=1,2,3$, and the DFs $P_2(t)$ and $P_3(t)$ increase to about 0.8, showing in this way a certain attitude of both these parties to form alliances.

 \begin{figure}
 			\begin{center}
 		\hspace*{-0.2cm}\includegraphics[width=8cm]{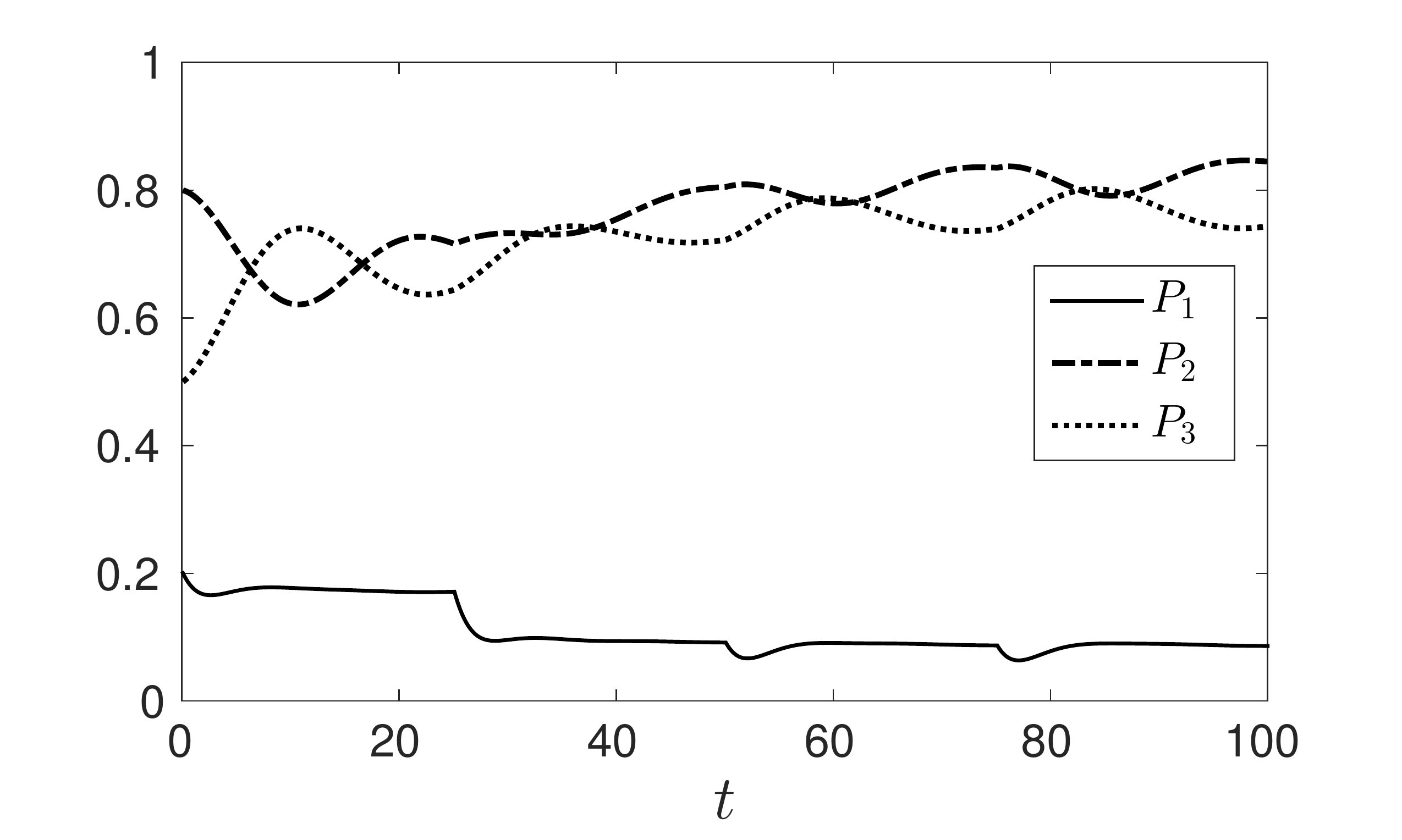}
 			\end{center}
 \vspace*{-0.5cm}\caption{ Initial conditions are: $n_{1,0}=0.2, n_{2,0}=0.8, n_{3,0}=0.5$, $N_{1,[0]}=0.1, N_{2,[0]}=0.6, N_{3,[0]}=1, N_{[0]}=1$. The choice of the parameters is the following:
 $\lambda_1=0.08,\lambda_2=\lambda_3=0.02, \tilde\lambda_1=0.01$ and $\nu^p_{12}=\nu^{ap}_{12}=\nu^{ap}_{13}=0.001$. The other parameters are set to 0.
 }
 \label{rulesFIG2}
 \end{figure}

 \subsubsection{}\label{sec:caserule3}
The third application proposed simulates a condition in which all the parties are listening, but not in the same way, to what all the electors are suggesting. Hence all the parameters are chosen different from 0, with the exception of the $\nu^{ap}_{k,l}$, which are all taken to be 0.
The initial conditions for this configuration are   $n_{1,0}=0.5, n_{2,0}=0.5, n_{3,0}=0.5$, $N_{1,[0]}=0.4, N_{2,[0]}=0.6, N_{3,[0]}=0.6, N_{[0]}=0.5$, and the other parameters are
$\omega_1=\omega_2=\omega_3= 0.01, \Omega_1=\Omega_2=\Omega_3=\Omega=0.1, \lambda_1=\lambda_3=0.001,
\lambda_2=0.02,\tilde\lambda_1=\tilde\lambda_2=0.01,\tilde\lambda_3=0.02$, and  $\nu^p_{k,l}=0.1$ for all $k,l$.
This choice  describes the fact that the parties $\Pc_1,\Pc_3$ are more influenced by the suggestion of the electors of the other parties rather than by their own electors, and this is the reason why the related  $\nu^p_{k,l}$ are larger than  $\lambda_1,\lambda_3$.
Hence, we expect that the role of the  electors $\Rc_1,\Rc_3$ is not so relevant, while the electors $\Rc_{und},\Rc_2$
should play an essential role in the time evolution of the DFs.

These DFs are shown in Fig. \ref{rulesFIG5}, and the previous analysis is confirmed. The function $P_1(t)$ increases in time, although  the electors $\Rc_1$ suggest not to ally (all the  $N_{1,[k]}$ are 0 after the application of the rule at each iteration). A similar behavior is observed for the party $\Pc_2$, but in this case the choice of the parameters
$\lambda_2=0.02$, which is bigger than the others, indicates that the party essentially follows $\Rc_2$. Regarding $\Pc_3$ we can observe that its DF is only slightly increasing in time, and it does not differ too much from the value 0.5. This is due to the choice $ \tilde\lambda_3=0.02$ that gives a significant role to the undecided electors $\Rc_{und}$ and to the fact that the various $N_{[k]}$ remain fixed at 0.5.

  \begin{figure}
  			\begin{center}
  		\hspace*{-0.2cm}\includegraphics[width=8cm]{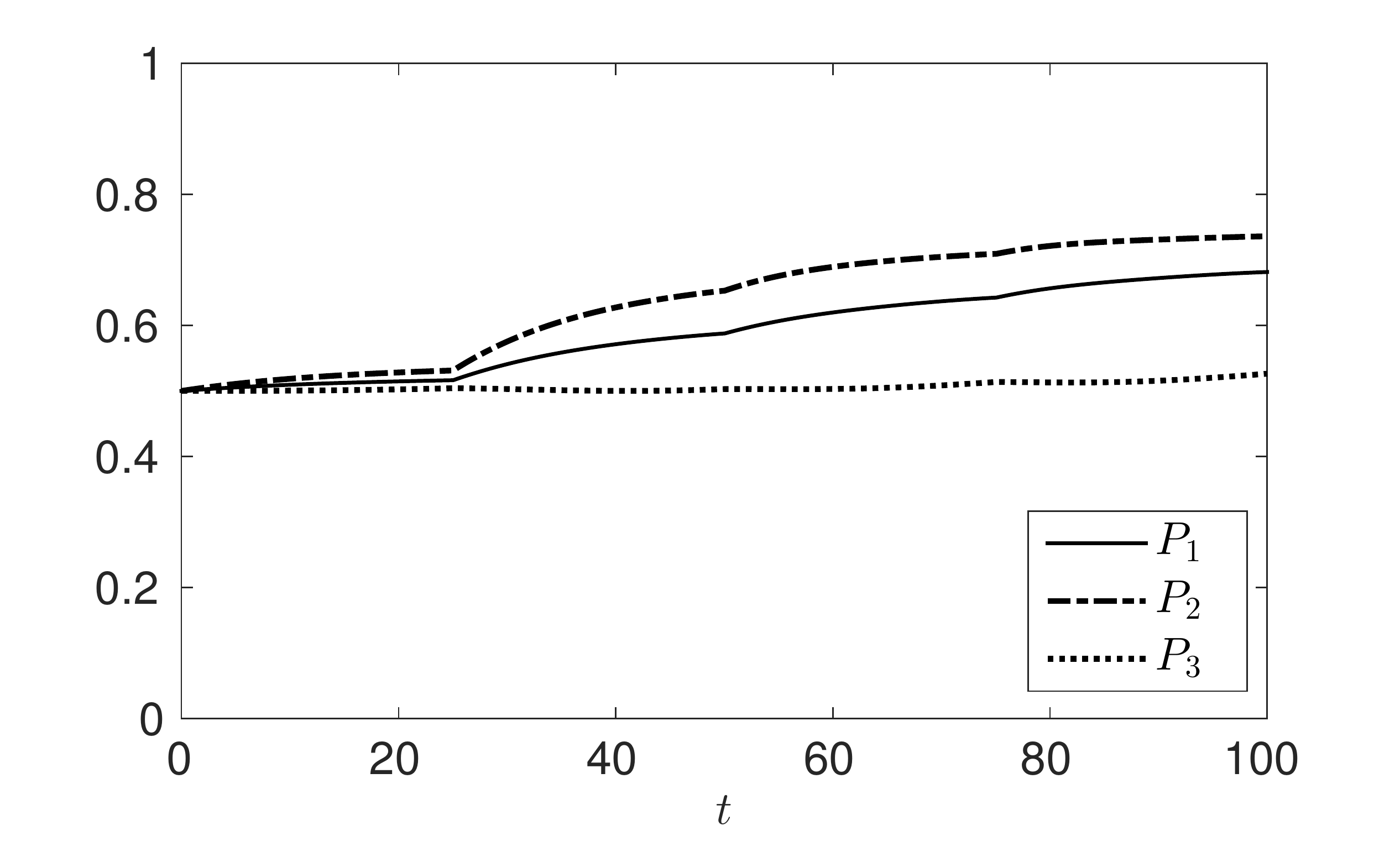}
  			\end{center}
  \vspace*{-0.5cm}\caption{Initial conditions are: $n_{1,0}=0.5, n_{2,0}=0.5, n_{3,0}=0.5$, $N_{1,[0]}=0.4, N_{2,[0]}=0.6, N_{3,[0]}=0.6, N_{[0]}=0.5$. The choice of the parameters is the following:
   $\omega_1=\omega_2=\omega_3= 0.01, \Omega_1=\Omega_2=\Omega_3=\Omega=0.1, \lambda_1=\lambda_3=0.001,
   \lambda_2=0.02,\tilde\lambda_1=\tilde\lambda_2=0.01,\tilde\lambda_3=0.02$, and  $\nu^p_{k,l}=0.1$ for all $k,l$. Other parameters are set to 0.}
  \label{rulesFIG5}
  \end{figure}

\section{The nonlinear model}\label{sectnonlinmod}
In this section we propose a somehow different model, without any rule and in which the parties do not interact with their electors, but only among themselves.
Hence, we neglect the effect of the reservoir in the Hamiltonian in \eqref{31}. On the other hand, we include some cubic and quartic interactions in the Hamiltonian, and we explore their consequences in the dynamics of the system. We can imagine that we are far away from elections, so that the parties are not particularly interested in what the electors have to say. For this reason, all the terms in $h$ involving interactions between the parties and the electors are neglected, and in fact the related parameters in $h$ are taken to be  zero. In \cite{all1} it was shown that, in presence of purely quadratic interactions, the DFs oscillates very much and no asymptotic limit (i.e., no final decision) can be reached. Here we will see that the DFs  still oscillate, but around certain asymptotic values which are very close to their initial values. This will be clear in the following.
In a previous work, \cite{all2}, the effects of cubic interactions were included in the Hamiltonian of the system, and the authors performed a perturbative analysis to recover an approximated solution. However, the model was so complicated that, even the perturbative approach was not efficient enough to produce any explicit solution. This was due to the presence of the reservoirs, which made all the computation rather un-friendly. Here, removing the reservoir, we are able to include different kind (cubic and quartic) of interactions in the Hamiltonian, and still to deduce exact solutions for the DFs of the parties.  Our new Hamiltonian $\hat H$ is the following:
\be
\left\{
\begin{array}{ll}
\hat H = H_{0}+H_{int}+H_{nl,1}+H_{nl,2},  \\
H_{0}=\sum_{j=1}^{3}\omega _{j}p_j^\dagger p_j,   \\
H_{int}=\mu_{12}^{ex}\left(p_1^\dagger p_2+p_2^\dagger p_1\right)+\mu_{12}^{coop}\left(p_1^\dagger p_2^\dagger+p_2 p_1\right)+\\
\qquad\mu_{13}^{ex}\left(p_1^\dagger p_3+p_3^\dagger p_1\right)  +\mu_{13}^{coop}\left(p_1^\dagger p_3^\dagger+p_3 p_1\right)+\\
\qquad\mu_{23}^{ex}\left(p_2^\dagger p_3+p_3^\dagger p_2\right)+\mu_{23}^{coop}\left(p_2^\dagger p_3^\dagger+p_3 p_2\right),\\
H_{nl,1}=\lambda^a_{12}(p_1^\dagger \hat P_2+\hat P_2 p_1)+
\lambda^a_{13}(p_1^\dagger \hat P_3+\hat P_3 p_1)+\lambda^a_{23}(p_2^\dagger \hat P_3+\hat P_3 p_2)+\\
\qquad \lambda^b_{12}(p_2^\dagger \hat P_1+\hat P_1 p_2)+\lambda^b_{13}(p_3^\dagger \hat P_1+\hat P_1 p_3)+\lambda^b_{23}(p_3^\dagger \hat P_2+\hat P_2 p_3),\\
H_{nl,2}=\tilde\mu_{12}^{ex}\hat P_3\left(p_1^\dagger p_2+p_2^\dagger p_1\right)+\tilde\mu_{12}^{coop}\hat P_3\left(p_1^\dagger p_2^\dagger+p_2 p_1\right)+\\
\qquad\tilde\mu_{13}^{ex}\hat P_2\left(p_1^\dagger p_3+p_3^\dagger p_1\right)  +\tilde\mu_{13}^{coop}\hat P_2\left(p_1^\dagger p_3^\dagger+p_3 p_1\right)+\\
\qquad\tilde\mu_{23}^{ex}\hat P_1\left(p_2^\dagger p_3+p_3^\dagger p_2\right)+\tilde\mu_{23}^{coop}\hat P_1\left(p_2^\dagger p_3^\dagger+p_3 p_2\right),
\end{array}%
\right.
\label{HamNL}\en
where $\omega _{j}$, $\lambda^{ab}_{ij}$, $\mu_{ij}^{ex},\tilde\mu_{ij}^{ex}$ and $\tilde\mu_{ij}^{coop}$ are real quantities, so that $\hat H=\hat H^\dagger$.

The meaning of the terms $H_0$ and $H_{int}$ was already explained in Section \ref{sec:themodel}.
The presence of the terms $H_{nl,1},H_{nl,2}$ introduces some new interactions between the parties. For instance the term $ \hat P_3 p_1^\dagger p_2^\dagger$ describes the fact that  $\Pc_1$ and $\Pc_2$ becomes more and more interested to form alliances for higher values of the DF of $\Pc_3$. Analogously, the term $p_1^\dagger \hat P_2$ describes the fact that $\Pc_1$ gets more interested in alliances the higher the attitude of $\Pc_2$ to ally is. And so on.
We say that the interactions are \textit{nonlinear} because the Heisenberg equations of motion deduced out of $\hat H$ in (\ref{HamNL}) are nonlinear. The interesting aspect is that $\hat H$ describes situations in which the three parties interact among them simultaneously, something which was not allowed by (\ref{31}), where we have only {\em two-bodies interactions}.

Because of these nonlinearities, it is technically convenient to work in the Schr\"odinger, rather than the Heisenberg, picture, see \cite{mer,rom,bagbook}. In this situation, calling $\Psi_0$ the vector describing the system at $t=0$, the decision function can be written as
\begin{equation}
P_j(t)=\langle\Psi(t),\hat P_j(0) \Psi(t) \rangle,
\label{DFnl}
\end{equation}
where $\Psi(t)=e^{-i\hat Ht}\Psi_0$ is the time evolution of $\Psi_0$.
Suppose now, as in Section \ref{sec:themodel}, that $\Psi_0=\sum\limits_{k,j,l=0}^{1}\alpha_{jkl}(0)\varphi_{k,j,l}$, then,  since $\Psi(t)=e^{-i\hat Ht} \Psi_0$ is the solution of the Schr\"{o}dinger equation
\begin{equation}
i\frac{\partial \Psi(t)}{\partial t}=\hat H\Psi(t),
\label{SHnl1}
\end{equation}
we get the following system of differential equations for the time-depending coefficients $\alpha_{k,j,l}(t)$ of $\Psi(t)$:
\begin{equation}
i\frac{\partial\alpha_{k,j,l}(t)}{\partial_t}=i\langle\varphi_{k,j,l},\frac{\partial \Psi}{\partial t}\rangle=\langle\varphi_{k,j,l},H\Psi(t)\rangle.
\label{SHnl2}
\end{equation}
The solution of this  system  returns the time evolution of the modes $\alpha_{k,j,l}(t)$, and the mean values $P_j(t)$ are finally deduced as in (\ref{add3})
\bea
\nonumber P_1(t)&=&\left<\hat P_1(t)\right> = \sum\limits_{j,l=0,1}|\alpha_{1,j,l}(t)|^2\\
\nonumber P_2(t)&=&\left<\hat P_2(t)\right>=
\sum\limits_{k,l=0,1}|\alpha_{k,1,l}(t)|^2\\
P_3(t)&=&\left<\hat P_3(t)\right>=\sum\limits_{k,j=0,1}|\alpha_{k,j,1}(t)|^2.
\label{DFsSH}
\ena

These formulas are used in the rest of this section to compute the DFs of the parties in some particular case, in order to highlight the various effects induced by $H_{nl,1}$ and $H_{nl,2}$ to the time evolution of the DFs.

\subsection{A first analytic analysis on the nonlinear term $H_{nl,1}$}\label{sec:qaHNL1}

\subsubsection{Competitive interactions}\label{sectcompint}
In order to get a better comprehension of the terms in $\hat H$, we first consider $H_{nl,2}=0$, and we assume that only competitive interactions
arise between the parties (i.e. $\mu^{coop}_{kj}=0$ for all $k,l$). Moreover, to simplify further the situation, we suppose that only  $\Pc_1$ and $\Pc_2$ interact. Hence, all the interaction parameters involving
the possible interactions of  $\Pc_3$ with $\Pc_1$ and $\Pc_2$ are put  to zero. We also consider $H_0=0$ here, since its role in the dynamics of a system is nowadays well understood, \cite{bagbook}, and it is related to the inertia of the various compartments of the model. In this way, we can reduce the number of free parameters used in the model. Summarizing, for the moment only  $\lambda^a_{12}$ and $\mu^{ex}_{12}$ are assumed to be different from zero.

The system of equations \eqref{SHnl2} can be written as
\begin{equation}
\dot{\Upsilon}(t)=-iB\Upsilon(t),
\label{SHnlcas1}
\end{equation}

where

$$
\begin{array}{cc}

\Upsilon(t)=\left(
       \begin{array}{c}
         \alpha_{0,0,0}(t) \\
         \alpha_{1,0,0}(t) \\
         \alpha_{0,1,0}(t) \\
         \alpha_{1,1,1}(t) \\
           \alpha_{0,0,1}(t) \\
                  \alpha_{1,0,1}(t) \\
                  \alpha_{0,1,1}(t) \\
                  \alpha_{1,1,1}(t) \\
       \end{array}
     \right), &
     B=\left(
     \begin{array}{cccccccc}
      0 & 0 & 0 & 0 & 0 & 0 & 0 & 0 \\
      0 & 0 & \mu^{ex}_{12} & 0 & 0 & 0 & 0 & 0 \\
      0 & \mu ^{ex}_{12} & 0 & \lambda^a_{12} & 0 & 0 & 0 & 0 \\
      0 & 0 & \lambda^a_{12} & 0 & 0 & 0 & 0 & 0 \\
      0 & 0 & 0 & 0 & 0 & 0 & 0 & 0 \\
      0 & 0 & 0 & 0 & 0 & 0 & \mu ^{ex}_{12} & 0 \\
      0 & 0 & 0 & 0 & 0 & \mu^{ex}_{12} & 0 & \lambda^a_{12} \\
      0 & 0 & 0 & 0 & 0 & 0 & \lambda^a_{12} & 0
     \end{array}
     \right).

\end{array}
$$

With this choice  it is not difficult to find an analytic solution of \eqref{SHnlcas1}, and
the DFs, deduced as in \eqref{DFsSH}, read as
\beano
  P_1(t)&=&\left(|\alpha_{1,0,0}(0)|^2+|\alpha_{1,0,1}(0)|^2\right)\left(\frac{(\lambda^a_{12})^2+(\mu^{ex}_{12})^2\cos^2(rt)}{(\lambda^a_{12})^2+(\mu^{ex}_{12})^2}\right)+\left(|\alpha_{0,1,0}(0)|^2+|\alpha_{0,1,1}(0)|^2\right)\sin^2(rt)+\\
 &&\left(|\alpha_{1,1,0}(0)|^2+|\alpha_{1,1,1}(0)|^2\right)\left(\frac{(\mu^{ex}_{12})^2+(\lambda^{a}_{12})^2\cos^2(rt)}{(\lambda^a_{12})^2+(\mu^{ex}_{12})^2}\right)+\\
 &&\left(\alpha_{1,0,0}(0)\alpha_{0,1,0}(0)+\alpha_{1,0,1}(0)\alpha_{0,1,1}(0)\right)\left(\frac{\mu^{ex}_{12}\lambda^{a}_{12}\sin^2(rt)}{(\lambda^a_{12})^2+(\mu^{ex}_{12})^2}\right),\\
\\
 P_2(t)&=&\left(|\alpha_{1,0,0}(0)|^2+|\alpha_{1,0,1}(0)|^2\right)\left(2\frac{(\lambda^a_{12}\mu^{ex}_{12})^2+(1+\cos(rt))(\mu^{ex}_{12})^4}{\left((\lambda^a_{12})^2+(\mu^{ex}_{12})^2\right)^2}\sin^2(rt)\right)+\\
&&\left(|\alpha_{0,1,0}(0)|^2+|\alpha_{0,1,1}(0)|^2\right)\left(\frac{(\lambda^a_{12})^2+(\mu^{ex}_{12})^2\cos^2(rt)}{(\lambda^a_{12})^2+(\mu^{ex}_{12})^2}\right)+\\
&&\left(|\alpha_{1,1,0}(0)|^2+|\alpha_{1,1,1}(0)|^2\right)\left(\frac{(\lambda^a_{12})^4+(\mu^{ex}_{12})^4+
(\lambda^a_{12}\mu^{ex}_{12})^2(-1-4\cos(rt)+\cos(2rt)}{\left((\lambda^a_{12})^2+(\mu^{ex}_{12})^2\right)^2}\right)+\\
&&\left(\alpha_{1,0,0}(0)\alpha_{0,1,0}(0)+\alpha_{1,0,1}(0)\alpha_{0,1,1}(0)\right)\left(\frac{-4\lambda^a_{12}\mu^{ex}_{12}\left((\lambda^a_{12})^2+(\mu^{ex}_{12})^2\cos^2(rt)\right)\sin^2(rt/2)}{\left((\lambda^a_{12})^2+(\mu^{ex}_{12})^2\right)^2}\right)\\
 P_3(t)&=& P_3(0),
\enano
where $r=\sqrt{(\lambda^a_{12})^2+(\mu^{ex}_{12})^2}$.

The first obvious remark is that $P_3(t)$ remains constant. This is expected, because of our choice of the parameters which excludes any interaction between $\Pc_3$ and the rest of the system. It is also possible to see that, if we increase the ratio $\lambda^a_{12}/\mu^{ex}_{12}$, then, when $t$ diverges, $P_2(t)$ tends toward the value
$$\left(|\alpha_{0,1,0}(0)|^2+|\alpha_{0,1,1}(0)|^2\right)+\left(|\alpha_{1,1,0}(0)|^2+|\alpha_{1,1,1}(0)|^2\right),$$
which coincides with $P_2(0)$. Moreover, also the amplitude of the oscillations of $P_2(t)$ depends on the same ratio $\lambda^a_{12}/\mu^{ex}_{12}$.
The existence of an asymptotic limit for $P_1(t)$ is not granted, in general, since it depends on the specific initial conditions.

If we now fix, in particular,  $\alpha_{1,0,0}=1$, whereas the other $\alpha_{k,j,l}$ are all zero, then
\beano
 P_1(t)&=&\frac{(\lambda^a_{12})^2+(\mu^{ex}_{12})^2\cos^2(rt)}{(\lambda^a_{12})^2+(\mu^{ex}_{12})^2},\\
 P_2(t)&=&2\,\frac{(\lambda^a_{12}\mu^{ex}_{12})^2+(1+\cos(rt))(\mu^{ex}_{12})^4}{\left((\lambda^a_{12})^2+
 (\mu^{ex}_{12})^2\right)^2}\sin^2(rt).
\enano
Hence both $P_1$ and $P_2$ oscillate with amplitudes that decrease if we increase the rate $\lambda^a_{12}/\mu^{ex}_{12}$. In particular
 as $\lambda^a_{12}/\mu^{ex}_{12}\rightarrow\infty$ then  $P_1(t)\rightarrow P_1(0)=1,P_2(t)\rightarrow P_2(0)=0$ for diverging $t$, so that both $P_1(t)$ and $P_2(t)$
 tend to their initial value. This  suggests that, adding the term $\lambda^{a}_{kj}\hat P_j\left(p_k^\dagger+p_k\right)$ in the Hamiltonian, forces the DF of $\Pc_j$ to stay close to its initial value, at least for sufficiently large values of $\lambda^a_{kj}/\mu^{ex}_{kj}$. In a certain sense, this term behaves as an {\em extended} inertial term, even if it is completely different from the inertia coming from  $H_0$, see \cite{bagbook}. It is now clear why we have taken $H_0=0$ in the present discussion: this was useful to avoid overlapping effects, which could have hidden this result.

\subsubsection{Cooperative interaction}

No particular difference arises if we now  suppose that  $\Pc_1$ and $\Pc_2$  have a cooperative attitude (i.e. $\mu^{coop}_{12}\neq0$). As before all the parameters regarding
the interactions of the party $\Pc_3$ with $\Pc_1$ and $\Pc_2$ are put equal to zero, and we further fix $\omega_1=\omega_2=0$. Moreover we now take $\mu^{ex}_{12}=0$. Hence $\lambda^a_{12}$ and $\mu^{coop}_{12}$ are the only parameters different from zero.

  In this case the matrix $B$ in \eqref{SHnlcas1} is
  $$
  B=\left(
       \begin{array}{cccccccc}
        0 & 0 & 0 & \mu ^{coop}_{12} & 0 & 0 & 0 & 0 \\
        0 & 0 & 0 & 0 & 0 & 0 & 0 & 0 \\
        0 & 0 & 0 & \lambda^a_{12} & 0 & 0 & 0 & 0 \\
        \mu ^{coop}_{12} & 0 & \lambda^a_{12} & 0 & 0 & 0 & 0 & 0 \\
        0 & 0 & 0 & 0 & 0 & 0 & 0 & \mu ^{coop}_{12} \\
        0 & 0 & 0 & 0 & 0 & 0 & 0 & 0 \\
        0 & 0 & 0 & 0 & 0 & 0 & 0 & \lambda^a_{12} \\
        0 & 0 & 0 & 0 & \mu ^{coop}_{12} & 0 & \lambda^a_{12} & 0
       \end{array}
       \right).
  $$

As before, it is not difficult to find an exact solution of the system in (\ref{SHnlcas1}). However, we just consider here what happens if we take $\alpha_{1,1,0}=1$, with all the other $\alpha_{k,j,l}$ equal to zero. Then, putting $\tilde r=\sqrt{(\lambda^a_{12})^2+(\mu^{coop}_{12})^2}$,
we find
\beano
   P_1(t)&=&\cos^2(\tilde rt),\\
   P_2(t)&=&\frac{(\lambda^a_{12})^2+(\mu^{coop}_{12})^2\cos^2(\tilde rt)}{(\lambda^a_{12})^2+(\mu^{coop}_{12})^2},
\enano
while $P_3(t)=P_3(0)$ for all $t$. Hence $P_1$ oscillates between  0 and 1, and $P_2$ oscillates between 1 and $\frac{(\lambda^a_{12})^2}{(\lambda^a_{12})^2+(\mu^{coop}_{12})^2}$. If we increase $\lambda^a_{12}/\mu^{coop}_{12}\rightarrow\infty$, then $P_2(t)\rightarrow P_0(t)=1$. This is the same effect observed in Section \ref{sectcompint}.

Also different choices of the initial status of the system suggest that, as in Section \ref{sectcompint},  $P_2(t)$ is confided to oscillate close to its initial value, with an amplitude of the oscillations decreasing when the ratio $\lambda^a_{12}/\mu^{coop}_{12}$ increases.
This confirms our previous claim that the term $\lambda^{a}_{kj}\hat P_j\left(p_k^\dagger+p_k\right)$ in the Hamiltonian behaves as a   { generalized inertia} for $\Pc_j$, since its effect for $P_j(t)$ is to produce oscillations of the DF for $\Pc_j$ around the initial value $P_j(0)$, with an amplitude which can be controlled.

\subsection{An analysis on the nonlinear term $H_{nl,2}$}

We begin this analysis by first taking $H_0=H_{nl,1}=0$ in (\ref{HamNL}). Moreover,
we suppose that  $\Pc_3$ can have competitive interactions
with $\Pc_1$ and $\Pc_2$ (i.e. $\mu^{ex}_{13}$ and $\mu^{ex}_{23}$ are non zero), while
$\Pc_1$ and $\Pc_2$ interact only if $\Pc_3$ is willing to ally
(i.e $\tilde\mu^{ex}_{12}\neq0,\tilde\mu^{coop}_{12}\neq0$).
In this case, taking also $\mu^{ex}_{13}=\mu^{ex}_{23}$, the matrix $B$ in \eqref{SHnlcas1} is
$$
  B=\left(
       \begin{array}{cccccccc}
        0 & 0 & 0 & 0 & 0 & 0 & 0 & 0 \\
        0 & 0 & 0 & 0 & \mu^{ex}_{13} & 0 & 0 & 0 \\
        0 & 0 & 0 & 0 &  \mu^{ex}_{13} & 0 & 0 & 0 \\
        0 & 0 & 0 & 0 & 0 &  \mu^{ex}_{13} & -\mu^{ex}_{13} & 0 \\
        0 & \mu^{ex}_{13} &  \mu^{ex}_{13} & 0 & 0 & 0 & 0 & \tilde\mu ^{coop}_{12} \\
        0 & 0 & 0 &  \mu^{ex}_{13} & 0 & 0 & \tilde\mu ^{ex}_{12} & 0 \\
        0 & 0 & 0 & -\mu^{ex}_{13} & 0 & \tilde\mu ^{ex}_{12} & 0 & 0 \\
        0 & 0 & 0 & 0 & \tilde\mu ^{coop}_{12} & 0 & 0 & 0
       \end{array}
       \right).
  $$

If we take the initial condition  $\alpha_{1,1,0}=1$ and the other $\alpha_{k,j,l}=0$, only the first two parties want initially to ally. By putting
$r_1=\sqrt{8(\mu^{ex}_{13})^2+(\tilde\mu^{ex}_{12})^2}$, the  DFs obtained are
\beano
P_1(t)&=& P_2(t) = \frac{1}{8({\mu}^{ex}_{13})^2+(\tilde{\mu}^{ex}_{12})^2}
 \left[ 6(\mu^{ex}_{13})^2+(\tilde\mu^{ex}_{12})^2+ 2(\mu^{ex}_{13})^2\cos(r_1t)) \right],\\
 P_3(t)&=&\frac{4\left(1-\cos(r_1t)\right)(\mu^{ex}_{12})^2}
 {8({\mu}^{ex}_{13})^2+(\tilde{\mu}^{ex}_{12})^2}.
\enano

Notice that in this case the DFs do not depend on the parameter $\tilde{\mu}^{coop}_{12}$.
It is clear that if we increase the ratio $\tilde{\mu}^{ex}_{12}/{\mu}^{ex}_{13}$ then $P_1(t),P_2(t)\rightarrow1$, while $P_3(t)\rightarrow0$.
Hence,  the DFs of the parties do no change too much from their initial values.

{This situation does not change if we consider as initial condition $\alpha_{1,0,0}=1$, and the other $\alpha_{k,j,l}=0$\footnote{Incidentally, let us notice that taking instead
$\alpha_{0,1,0}=1$ will produce the same solution but $P_1$ and $P_2$ are switched.}. In this case, calling $r_2=\sqrt{2(\mu^{ex}_{13})^2+(\tilde\mu^{coop}_{12})^2}$, we obtain

\beano
P_1(t)&=&\frac{1}{\left(2({\mu}^{ex}_{13})^2+(\tilde{\mu}^{coop}_{12})^2\right)^4}\left[
   (\tilde\mu^{coop}_{12})^8+7(\tilde\mu^{coop}_{12})^6(\mu^{ex}_{13})^8+
   21(\tilde\mu^{coop}_{12})^4(\mu^{ex}_13)^4+16(\tilde\mu^{coop}_{12})^2(\mu^{ex}_{13})^6+4(\mu^{coop}_{12})^8+\right.\\
&& \cos(r_2t)\left(8 (\mu^{ex}_{13})^8-14(\mu^{ex}_{13})^4(\tilde\mu^{coop}_{12})^6-2(\mu^{ex}_{13})^2(\tilde\mu^{coop}_{12})^6\right)+\\
 && \left. \cos^2(r_2t)\left(4(\tilde\mu^{coop}_{12})^2(\mu^{ex}_{13})^4+8(\mu^{ex}_{13})^4(\tilde\mu^{coop}_{12})^4+(\mu^{ex}_{13})^4(\tilde\mu^{coop}_{12})^6\right)  \right],\\
P_2(t)&=&\frac{1}{\left(2({\mu}^{ex}_{13})^2+(\tilde{\mu}^{coop}_{12})^2\right)^4}
 \left[ 4(\mu^{ex}_{13})^8+12(\mu^{ex}_{13})^6(\tilde\mu^{coop}_{12})^2+17(\mu^{ex}_{13})^4(\tilde\mu^{coop}_{12})^4+2
 (\mu^{ex}_{12})^2(\tilde\mu^{coop}_{12})^6 +\right.\\
 && -2\cos(r_2t)\left(4(\mu^{ex}_{13})^8+8(\mu^{ex}_{13})^6(\tilde\mu^{coop}_{12})^2+9(\mu^{ex}_{13})^4(\tilde\mu^{coop}_{12})^4+
 (\mu^{ex}_{13})^2(\tilde\mu^{coop}_{12})^6\right)+\\
&&\left. \cos^2(r_2t)\left(4(\tilde\mu^{coop}_{12})^2(\mu^{ex}_{13})^4+8(\mu^{ex}_{13})^4(\tilde\mu^{coop}_{12})^4+(\mu^{ex}_{13})^4(\tilde\mu^{coop}_{12})^6\right)  \right],\\
P_3(t)&=&\frac{2(\mu^{ex}_{13})^2\sin^2(r_2t)\left((\mu^{ex}_{13})^2+(\tilde{\mu}^{coop}_{12})^2+(\mu^{ex}_{13})^2\cos(r_2t)\right)}
{\left(2(\mu^{ex}_{13})^2+(\tilde{\mu}^{coop}_{12})^2\right)^2}
\enano

The DFs do not depend on the parameter $\tilde{\mu}^{ex}_{12}$. If we increase the rate $\tilde{\mu}^{coop}_{12}/{\mu}^{ex}_{13}$ then $P_1(t)\rightarrow1$, and $P_2(t),P_3(t)\rightarrow0$ and again, the DFs of the parties do not change very much from their initial value.}

A somehow different situation arises if we  take the initial condition  $\alpha_{1,0,1}=1$ and the other $\alpha_{k,j,l}=0$.
In this case we get
 \beano
 P_1(t)&=&\frac{1}{2\left( 8(\mu^{ex}_{13})^2+(\tilde\mu^{ex}_{12})^2 \right)}
 \left[
 \left(12(\mu^{ex}_{13})^2+2(\tilde\mu^{ex}_{12})^2+
 4 (\mu^{ex}_{14})^2\cos\left(r_1t\right)+\right.\right. \\
 && \left(8(\mu^{ex}_{13})^2+(\tilde\mu^{ex}_{12})^2-\tilde\mu^{ex}_{12}r_1 \right)\cos\left(t(-3(\tilde\mu^{ex}_{12}+r_1)/2\right) +\\
&&\left.  \left(8(\mu^{ex}_{13})^2+(\tilde\mu^{ex}_{12})^2+\tilde\mu^{ex}_{12}r_1\right)\cos\left(t(3(\tilde\mu^{ex}_{12}+r_1)/2\right) \right],\\
P_2(t)&=&\frac{1}{2\left( 8(\mu^{ex}_{13})^2+(\tilde\mu^{ex}_{12})^2 \right)}
 \left[
 \left(20(\mu^{ex}_{13})^2+2(\tilde\mu^{ex}_{12})^2-
 4 (\mu^{ex}_{13})^2\cos\left(r_1t\right)+\right.\right. \\
 && -\left(8(\mu^{ex}_{13})^2+(\tilde\mu^{ex}_{12})^2-\tilde\mu^{ex}_{12}r_1 \right)\cos\left(t(-3(\tilde\mu^{ex}_{12}+r_1)/2\right) +\\
&&\left.  -\left(8(\mu^{ex}_{13})^2+(\tilde\mu^{ex}_{12})^2+\tilde\mu^{ex}_{12}r_1)\right)
\cos\left(t(3(\tilde\mu^{ex}_{12}+r_1)/2\right) \right],\\
P_3(t)&=&\frac{1}{\left( 8(\mu^{ex}_{13})^2+(\tilde\mu^{ex}_{12})^2 \right)}\left(6(\mu^{ex}_{13})^2+(\tilde\mu^{ex}_{12})^2+2(\mu^{ex}_{13})^2\cos\left(r_1t\right) \right)
 \enano

We see that $P_1(t)$ and $P_2(t)$ both oscillate between 0 and 1, no matter the rate $\tilde{\mu}^{ex}_{12}/{\mu}^{ex}_{12}$ is large.
 On the other hand if $\tilde{\mu}^{ex}_{12}/{\mu}^{ex}_{12}$ increases, then $P_3(t)\rightarrow1$, which means that only $P_3(t)$ stays close to its initial value.

{Hence, adding the term $H_{nl,2}$ in the Hamiltonian leads to some dynamics already observed
in the analysis of the term $H_{nl,1}$: in fact, each term
$\tilde\mu_{kl}^{ex}\hat P_j\left(p_k^\dagger p_l+p_l^\dagger p_k\right)+\tilde\mu_{kl}^{coop}\hat P_j\left(p_k^\dagger p_l^\dagger+p_l p_k\right)$
has the effect to confine $P_j(t)$ close to its initial value if we increase the values $\tilde\mu_{kl}^{ex},\tilde\mu_{kl}^{coop}$ with respect to the other parameters. However, while  $H_{nl,1}$ describes interactions between  two parties,
$H_{nl,2}$ describes simultaneous interactions between the three parties.}

\subsubsection{More considerations for the nonlinear model}
 We end the analysis of this model by briefly discussing {three}  more scenarios arising from the Hamiltonian \eqref{HamNL}.

 In the first scenario we consider the case in which  $\Pc_1$ does not want to ally. Moreover $\Pc_1$ is not inclined to cooperate with the other parties, whereas it has an elevate competitive behavior. This means that we have to take the parameters
 $\lambda^b_{12},\lambda^b_{13},\tilde{\mu}^{ex}_{12},\tilde{\mu}^{coop}_{12},\mu^{ex}_{12,13}$ quite large, while $\mu^{coop}_{12}$ and $\mu^{coop}_{13}$ are quite low. On the other side, we assume that both $\Pc_2$ and $\Pc_3$ want to ally, and moreover they are inclined to cooperate, and they tend to be not  much competitive between them. Then we take $\tilde{\mu}^{ex}_{23},\tilde{\mu}^{coop}_{23},\mu^{ex}_{23}$  low. The difference between
 $\Pc_2$ and $\Pc_3$ relies in the fact that $\Pc_2$ is more \textit{static}, so we have to take  $\omega_2$ larger than $\omega_3$\footnote{This incidentally shows that we are here considering $H_0\neq0$.}.

 In Fig. \ref{NonLin1} we show the time evolution of the DFs, with the initial conditions $P_1(0)=0,P_2(0)=1,P_3(0)=1$, or, written adopting the notation used in system
 \eqref{SHnlcas1},  $\alpha_{0,1,1}=1$ and other $\alpha_{k,j,l}=0$. The parameters are chosen as follows:
 $\omega_1=0.3,\omega_2=0.5,\omega_3=0.1$, $\mu^{ex}_{12}=0.01,\mu^{ex}_{13}=\mu^{ex}_{23}=0.05$,
 $\mu^{coop}_{12}=0.05,\mu^{coop}_{13}=\mu^{coop}_{23}=0.2$,  $\lambda^{a}_{12}=0.03,\lambda^{a}_{13}=\lambda^{a}_{23}=0.1$,
 $\lambda^{b}_{12}=0.3,\lambda^{b}_{13}=\lambda^{b}_{23}=0.1$,
  $\tilde\mu^{ex}_{12}=0.03,\tilde\mu^{ex}_{13}=0.02,\tilde\mu^{ex}_{23}=0.3$,
  $\tilde\mu^{coop}_{12}=0.03,\tilde\mu^{coop}_{13}=0.02,\tilde\mu^{coop}_{23}=0.3$. The evolution shows that the attitude of $\Pc_1$ to  ally is always quite low, while $\Pc_2$ and $\Pc_3$ are more inclined to form alliances, with
 $\Pc_2$ even  more inclined than $\Pc_3$. This is also due to the higher value of the inertia of $\Pc_2$ compared to that of $\Pc_3$ (recall that $\omega_2>\omega_3$), which contributes to make smaller the amplitude of the oscillations of $P_2(t)$ with respect to those of $P_3(t)$.
 \begin{figure}[!]
 			\begin{center}
 		\hspace*{-0.2cm}\includegraphics[width=8cm]{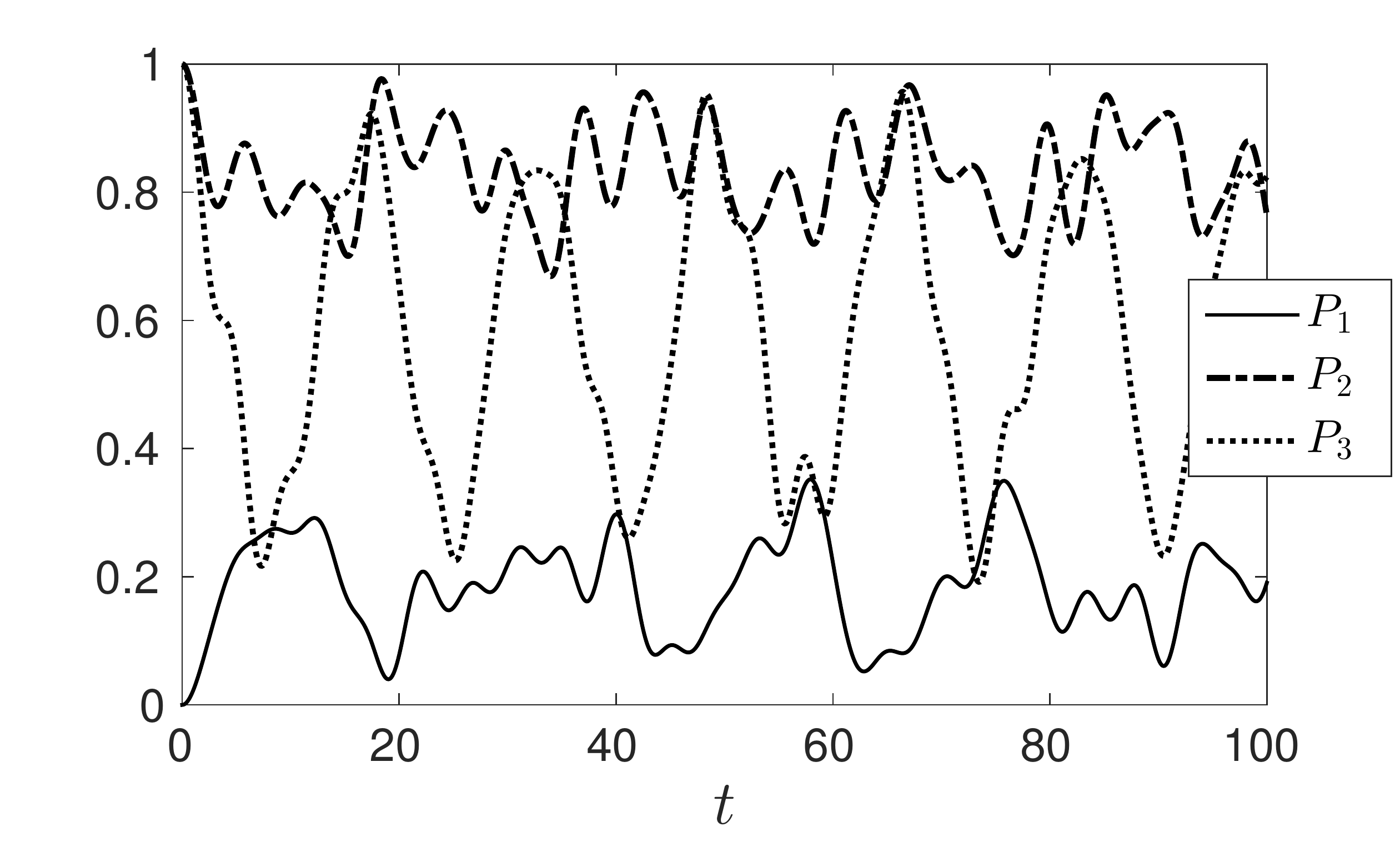}
 			\end{center}
 \vspace*{-0.5cm}\caption{Simulation of the nonlinear model. The initial conditions are  $\alpha_{0,1,1}=1$, while the other $\alpha_{k,j,l}=0$. The parameters are:
  $\omega_1=0.3,\omega_2=0.5,\omega_3=0.1$, $\mu^{ex}_{12}=0.01,\mu^{ex}_{13}=\mu^{ex}_{23}=0.05$,
  $\mu^{coop}_{12}=0.05,\mu^{coop}_{13}=\mu^{coop}_{23}=0.2$,  $\lambda^{a}_{12}=0.03,\lambda^{a}_{13}=\lambda^{a}_{23}=0.1$,
  $\lambda^{b}_{12}=0.3,\lambda^{b}_{13}=\lambda^{b}_{23}=0.1$,
   $\tilde\mu^{ex}_{12}=0.03,\tilde\mu^{ex}_{13}=0.02,\tilde\mu^{ex}_{23}=0.3$,
   $\tilde\mu^{coop}_{12}=0.03,\tilde\mu^{coop}_{13}=0.02,\tilde\mu^{coop}_{23}=0.3$. }
 \label{NonLin1}
 \end{figure}

In the second situation we consider the case in which $\Pc_1$ and $\Pc_3$ do not want to ally at $t=0$, whereas $\Pc_2$ is inclined to alliance. Hence $\alpha_{0,1,0}=1$, while the other $\alpha_{k,j,l}$ are zero. We consider the following choice of the parameters: $\omega_1=0.1,\omega_2=0.02,\omega_3=0.1$, $\mu^{ex}_{12}=\mu^{ex}_{13}=\mu^{ex}_{23}=0.05$,
 $\mu^{coop}_{12}=\mu^{coop}_{13}=\mu^{coop}_{23}=0.15$,  $\lambda^{a}_{12}=0.03,\lambda^{a}_{13}=\lambda^{a}_{23}=0.2$,
 $\lambda^{b}_{12}=\lambda^{b}_{13}=\lambda^{b}_{23}=0.2$,
  $\tilde\mu^{ex}_{12}=0.2,\tilde\mu^{ex}_{13}=0.02,\tilde\mu^{ex}_{23}=0.5$,
  $\tilde\mu^{coop}_{12}=0.2,\tilde\mu^{coop}_{13}=0.02,\tilde\mu^{coop}_{23}=0.5$.
This choice reflects the fact that, in particular, all the parties are quite competitive.   Results in Fig. \ref{NonLin2} shows that  $P_1(t)$ remains always quit low, while $P_2(t)$ and $P_3(t)$ have an higher variance.

  \begin{figure}
  			\begin{center}
  		\hspace*{-0.2cm}\includegraphics[width=8cm]{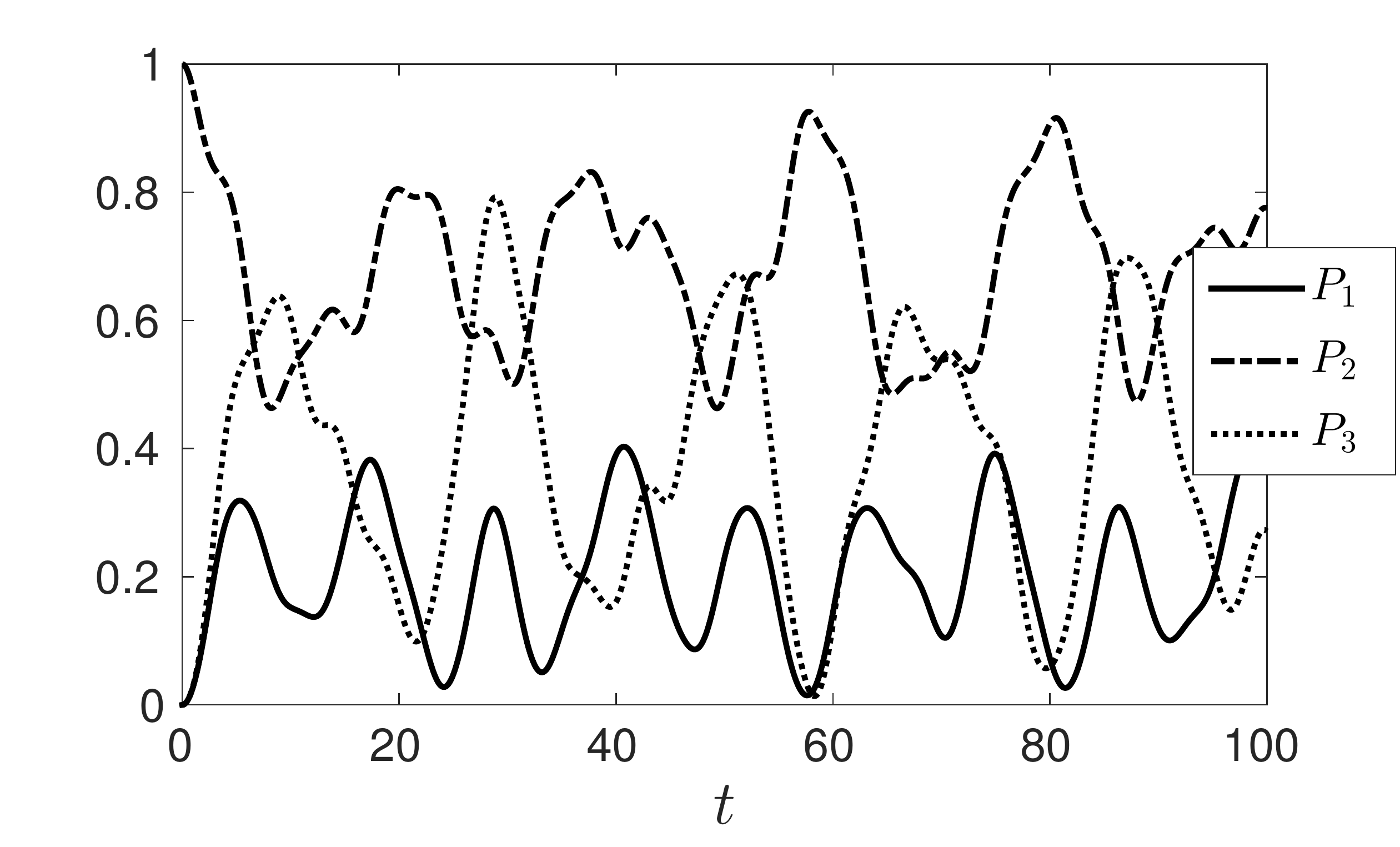}
  			\end{center}
  \vspace*{-0.5cm}\caption{The initial conditions are: $\alpha_{0,1,0}=1$, while the other $\alpha_{k,j,l}=0$. The parameters are:
    $\omega_1=0.15,\omega_2=0.15,\omega_3=0.15$, $\mu^{ex}_{12}=\mu^{ex}_{13}=\mu^{ex}_{23}=0.05$,
     $\mu^{coop}_{12}=\mu^{coop}_{13}=\mu^{coop}_{23}=0.05$,  $\lambda^{a}_{12}=0.03,\lambda^{a}_{13}=\lambda^{a}_{23}=0.2$,
     $\lambda^{b}_{12}=\lambda^{b}_{13}=\lambda^{b}_{23}=0.2$,
      $\tilde\mu^{ex}_{12}=0.2,\tilde\mu^{ex}_{13}=0.02,\tilde\mu^{ex}_{23}=0.5$,
      $\tilde\mu^{coop}_{12}=0.2,\tilde\mu^{coop}_{13}=0.02,\tilde\mu^{coop}_{23}=0.5$.  }
  \label{NonLin2}
  \end{figure}

 {Finally, we briefly sketch what happens changing the parameter $\lambda^{a}_{12}$ which
 tunes the strength of the nonlinear interaction given by the Hamiltonian term $H_{nl,1}$.
 We set the parameters as follows:
 $\omega_1=\omega_2=0.1,\omega_3=0.3$, $\mu^{ex}_{12}=\mu^{ex}_{13}=\mu^{ex}_{23}=0.05$,
  $\mu^{coop}_{12}=0.05,\mu^{coop}_{13}=\mu^{coop}_{23}=0.05$,  $\lambda^{a}_{13}=\lambda^{a}_{23}=0.2$,
  $\lambda^{b}_{12}=\lambda^{b}_{13}=\lambda^{b}_{23}=0.1$,
   $\tilde\mu^{ex}_{12}=\tilde\mu^{ex}_{13}=\tilde\mu^{ex}_{23}=0.1$,
   $\tilde\mu^{coop}_{12}=\tilde\mu^{coop}_{13}=\tilde\mu^{coop}_{23}=0.1$. The parameter  $\lambda^{a}_{12}$ varies between 0.5 and 2. Initial conditions are chosen   as $P_1(0)=0.6, P_2(0)=0.3, P_3(0)=0$ (or in terms of the $\alpha_{j,k,l}$, $\alpha_{1,0,0}=\sqrt{0.6},
    \alpha_{0,1,0}=\sqrt{0.3}, \alpha_{0,0,0}=\sqrt{0.1}$). The results are shown in Fig. \ref{Nonlin_lambda}. As expected, if we increase $\lambda^a_{12}$
    then $P_2(t)$  oscillates close to its initial value $P_2(0)=0.3$ with decreasing amplitude and increasing frequency. Conversely $P_1(t)$ experiences  high-frequency oscillations for increasing $\lambda^{a}_{12}$: this was justified in the qualitative analysis performed in Section \ref{sec:qaHNL1}, where it was shown, for some particular initial conditions, that $P_1(t)$ can develop high-frequency oscillations by increasing the rates $\lambda^{a}_{12}/\mu^{ex}_{12}$ and $\lambda^{a}_{12}/\mu^{coop}_{12}$. Regarding $P_3(t)$, we can  notice that it also experiences rapid oscillations for higher $\lambda^{a}_{12}$, although less marked than in the case of $P_1(t)$: this is because $\Pc_3$ interacts with $\Pc_1$, so that the oscillations in $P_1(t)$ are somewhat reflected by the behavior of $P_3(t)$.

 \begin{figure}[!]
 			\begin{center}
 		 \subfigure[$P_{1}(t)$]{\hspace*{-0.1cm}\includegraphics[width=7cm]{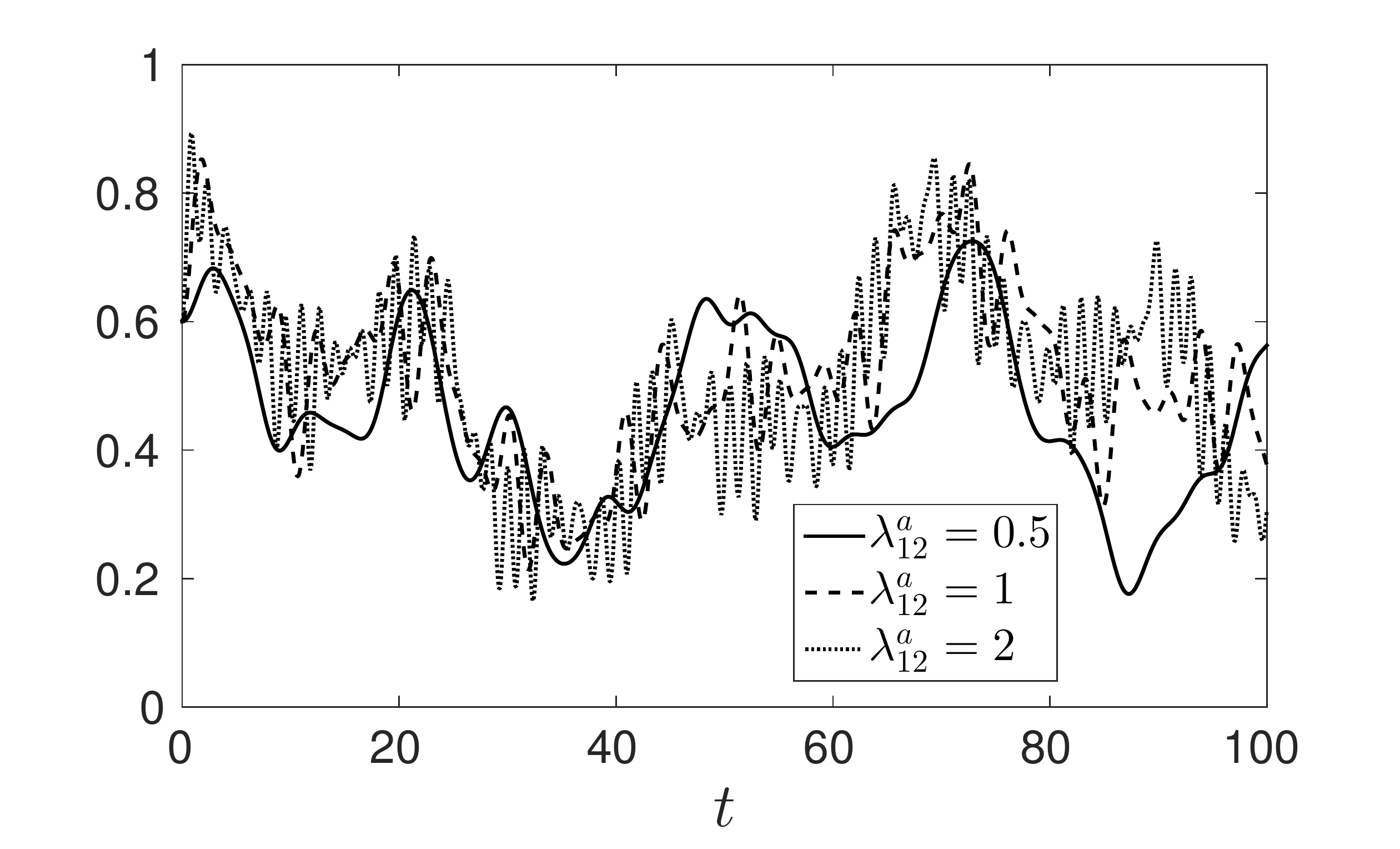}}
 		\subfigure[$P_{2}(t)$]{\hspace*{-0.1cm}\includegraphics[width=7cm]{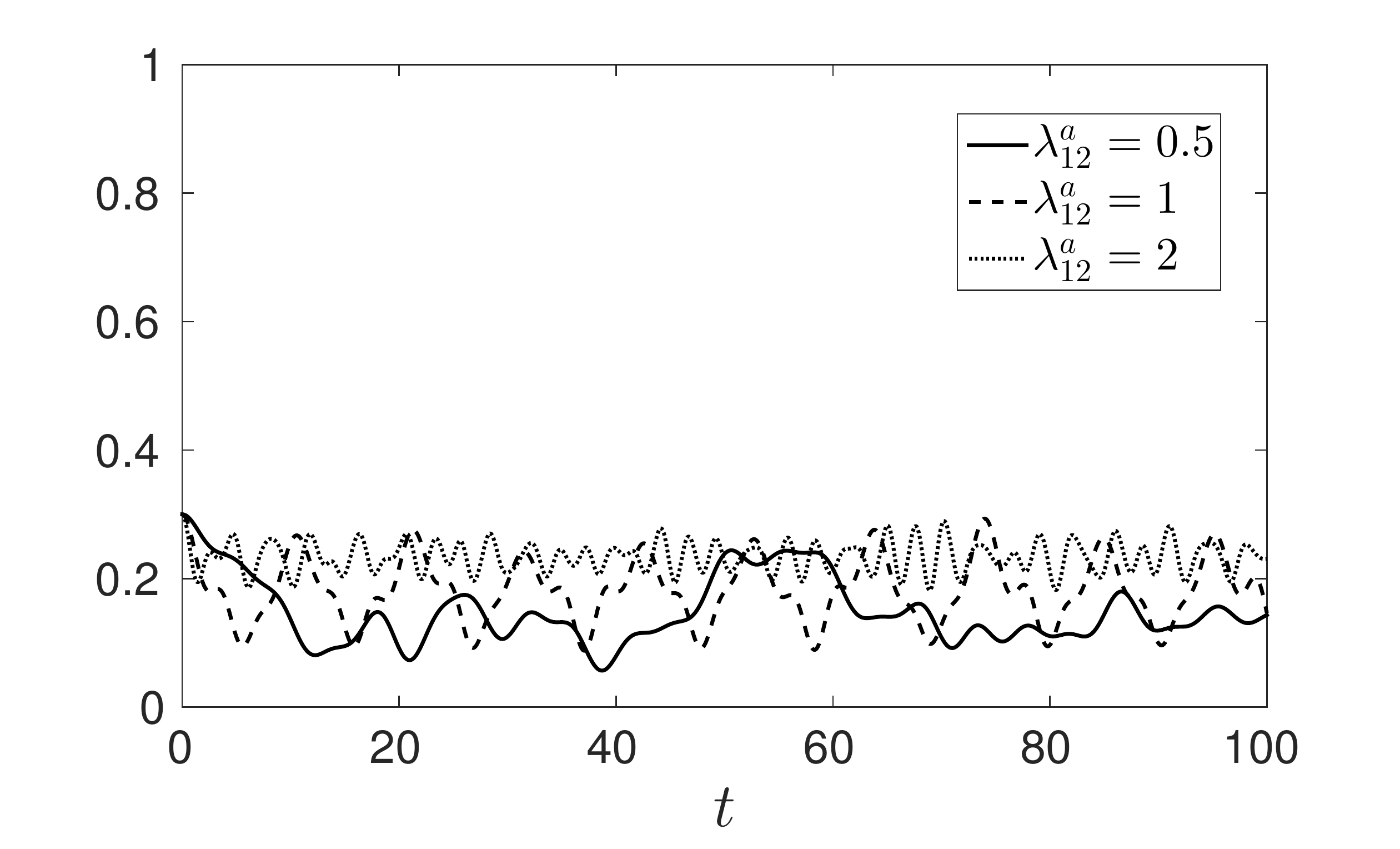}}	
 		\subfigure[$P_{3}(t)$]{\hspace*{-0.1cm}\includegraphics[width=7cm]{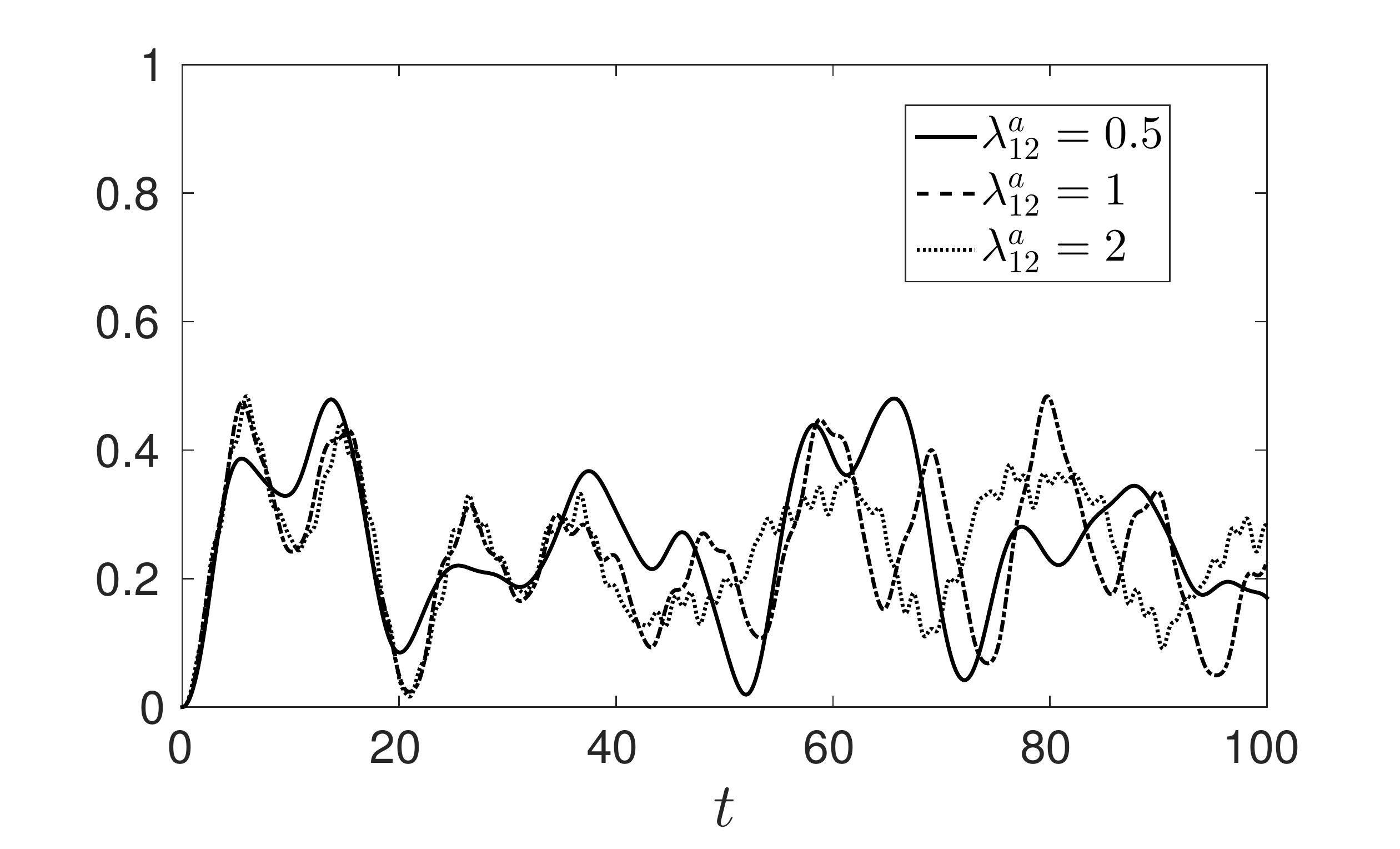}}	
 			\end{center}
 \vspace*{-0.5cm}\caption{Time evolution of the DFs for various values of $\lambda^{a}_{12}$. Initial conditions are $P_1(0)=0.6, P_2(0)=0.3, P_3(0)=0$. The other parameters are $\omega_1=\omega_2=0.1,\omega_3=0.3$, $\mu^{ex}_{12}=\mu^{ex}_{13}=\mu^{ex}_{23}=0.05$,
   $\mu^{coop}_{12}=0.05,\mu^{coop}_{13}=\mu^{coop}_{23}=0.05$,  $\lambda^{a}_{13}=\lambda^{a}_{23}=0.2$,
   $\lambda^{b}_{12}=\lambda^{b}_{13}=\lambda^{b}_{23}=0.1$,
    $\tilde\mu^{ex}_{12}=\tilde\mu^{ex}_{13}=\tilde\mu^{ex}_{23}=0.1$,
    $\tilde\mu^{coop}_{12}=\tilde\mu^{coop}_{13}=\tilde\mu^{coop}_{23}=0.1$.}
 \label{Nonlin_lambda}
 \end{figure}
}

\section{Conclusions and perspectives}\label{sectconcl}

In this paper we have proposed three different, but related, dynamical systems, all based on an operatorial approach, describing a political system with three parties, interacting or not with a set of supporters of different nature. In particular, the first model is based on an open system whose dynamics is driven (only) by a self-adjoint Hamiltonian $h$. The second model is a variation of this first model, since the dynamics is driven by the same $h$, but is also subjected to a certain rule implementing more relations between the supporters and the parties, relations which cannot be expressed directly in an Hamiltonian language. The third model, which can also be seen as a variation on the same theme, is based on the assumption that the interactions between the parties, if these are not purely quadratic, become more relevant than the interactions  between the parties and the supporters, which can then be neglected, at a first approximation. We believe that all these models have their own relevance, since they describe, with an unifying language, different, but related, situations. For instance, the presence of the rule allow us to include in the general settings the presence of some information, while using the non linear model proposed in Section\label{sectnonlinmod} we can test the effects of different interactions between the parties.

{ An interesting feature, evident when comparing the models considered in this paper, is that the nonlinearities considered in Section \ref{sectnonlinmod} can be efficiently  used to get a sort of (slightly oscillating) asymptotic value for each DF, avoiding in this way the use of any reservoir, with or without any rule. From the point of view of the interpretation, the model in Section \ref{sectnonlinmod} can be considered useful  when the elections are still too distant so that the parties do not care much about what the electors suggest.} Also, this suggests that  stronger (i.e., not purely quadratic) interactions between the parties can be useful to help the parties to reach a (almost) stable decision.

However,  we believe that the ones with the reservoir, and in particular the one with the rule proposed in Section \ref{sectrule}, is the most realistic within the models considered in this paper, since it takes into account several reasonable effects which are observed in real life  in many political systems. Also, in spite of its apparent difficulty, the time evolution of each DF can be deduced analytically.

Of course our model can be improved in several ways. For instance, we could consider effects due to different behaviors of different members of the parties, considering their specific communicative skills, their tendencies to be corrupted (effects already taken into account in \cite{FW94,BO14}), and more effects.  Another possible interesting generalization consists in adding some random effect in the models, possibly adding some noise to the parameters of the Hamiltonians (\ref{31}) or (\ref{HamNL}). In our opinion this approach could be used to model some irrational behavior of the electors.
Moreover we guess our model can be extended to a more general framework, including
democratic hierarchical bottom-up voting (see  the interesting works \cite{galam1,galam2,pol3}), or the evolution of the opinion in a closed group with respect to a choice between multiple options (\cite{RM05}).
All these possible generalizations would probably improve the reliability of our models, which, however, already at the present stage produce, we believe, a quite rich and interesting dynamics.

\renewcommand{\theequation}{A.\arabic{equation}}

\section*{Appendix:  Introducing the $(H,\rho)$-induced dynamics}

In a recent paper, \cite{BDGO},  the idea of $(H,\rho)$--induced dynamics was introduced, by putting together  the general framework of the quantum dynamics, described by a self-adjoint Hamiltonian $H$, and some periodic (or not) effect which cannot be included in $H$. This appendix is devoted to give some essential information on this  dynamics.

Let $\Sc$ be our physical system and let $O_j$ be a set of $M$ commuting self-adjoint operators, needed for the complete description of $\Sc$, with eigenvectors $\varphi^{(j)}_{\alpha_n}$ and eigenvalues $\alpha_n^{(j)}$:
\be
[O_j,O_k]=0, \qquad O_j=O_j^\dagger,\qquad  O_j\varphi^{(j)}_{n_j}=\alpha_{n_j}^{(j)}\varphi^{(j)}_{n_{j}},
\label{21}
\en
$j=1,2,\ldots,M$, $n_j=1,2,3,\ldots,N_j$, which can be finite or infinite.
Let $\mathbf{n}=(n_1,n_2,\ldots,n_M)$ and let
$$
\varphi_{\mathbf{n}}=\varphi^{(1)}_{n_{1}}\otimes\varphi^{(2)}_{n_{2}}\otimes\cdots\otimes\varphi^{(M)}_{n_{M}}.
$$
Then $\varphi_{\mathbf{n}}$ is an eigenstate of all the operators $O_j$, \emph{i.e.},
\be
O_j\,\varphi_{\mathbf{n}}=\alpha_{n_j}^{(j)}\,\varphi_{\mathbf{n}}.
\label{22app}
\en
It is convenient and natural to assume that these vectors are mutually orthogonal and normalized:
\be
\left<\varphi_{\mathbf{n}},\varphi_{\mathbf{m}}\right>=\delta_{\mathbf{n},\mathbf{m}}=\prod_{j=1}^M\delta_{n_j,m_j}.
\label{23app}
\en
The Hilbert space $\Hil$ where $\Sc$ is  defined is (mathematically) constructed as the closure of the linear span of all the vectors $\varphi_\mathbf{n}$, which therefore turn out to form an orthonormal basis for $\Hil$. Now, let $H=H^\dagger$ be the (time-independent) self-adjoint Hamiltonian of $\Sc$. This means that, in absence of any other information, the wave function $\Psi(t)$ describing $\Sc$ at time $t$ evolves according to the Schr\"odinger equation $i\dot\Psi(t)=H\Psi(t)$, where $\Psi(0)=\Psi_0$ describes the initial status of $\Sc$. The formal\footnote{The reason why we speak about a formal solution is that $\exp(-iHt)\Psi_0$ is not, in general, explicitly known, at least if there is no easy way to compute the action of the unitary operator $\exp(-iHt)$ on the vector $\Psi_0$, which is not granted at all. This is not very different from the equivalence of a differential equation with some given initial conditions and its integral counterpart: they contain the same information but none of them provide the explicit solution of the dynamical problem.} solution of the
Schr\"odinger equation in $t\in[0,\tau[$, for a fixed $\tau>0$, is
$\Psi_0(t)=\exp(-iHt)\Psi_0$. Let now $\rho$ be our {\em rule}, \emph{i.e.}, a set of conditions mapping, at a certain time, any input vector $\Psi_{0}(\tau)\in \Hil$ in a new
vector $\Psi_{1}\in\Hil$, and with a synthetic notation we will simply write $\Psi_{1}=\rho(\Psi_{0}(\tau))$. This is not very different from what happens in scattering theory, where an incoming state, after the occurrence of the scattering, is transformed into an outgoing state.
Then, in the new time interval $t\in[0,\tau[$, the new vector $\Psi_{1}$ evolves
according to the Schr\"odinger evolution, $\Psi_{1}(t)=\exp(-iHt)\Psi_{1}$, and at time $\tau$ we map $\Psi_{1}(\tau)$ into a new state $\Psi_2$.
The procedure can continue for more iterations $k=1,2,\cdots$. Let now $X$ be a generic operator on $\Hil$, bounded or unbounded. In this last case, we will require that the various $\Psi_k$  belong to the domain of $X(t)=\exp(iHt)X\exp(-iHt)$ for all $t\in[0,\tau]$. For instance in Section \ref{sectrule} we have considered $X$ as the number operators related to the various parties.

\begin{defn}\label{def1}
The sequence of functions
\be
x_{k}(t):=\left<\Psi_k, X(t)\Psi_k\right>,
\label{24}\en
for $t\in[0,\tau]$ and $k\in {\Bbb N}_0$, is called the $(H,\rho)$--induced dynamics of the operator $X$.
\end{defn}
We refer to \cite{BDGO} for more details of the $(H,\rho)$--induced dynamics. Here we just observe  that, using ${\underline X}(t)=(x_1(t),x_2(t),x_3(t),\ldots)$, it is possible to define a function of time in the following way:
\be
\tilde X(t)=
\left\{
\begin{array}{ll}
x_1(t),\qquad t\in [0,\tau[ &  \\
x_2(t-\tau),\qquad t\in [\tau,2\tau[ &  \\
x_3(t-2\tau),\qquad t\in [2\tau,3\tau[ &  \\
\ldots &
\end{array}%
\right.
\label{25bis}
\en
It is clear that $\tilde X(t)$ may have discontinuities in $k\tau$, for $k\in \Bbb N$.  In \cite{BDGO} we have discussed conditions for $\tilde X(t)$ to admit some asymptotic value or to be periodic, and some related notions of {\em equilibria} for a $(H,\rho)$-dynamics.



\begin{thebibliography}{99}



\bibitem{qdm4} M. Asano, M. Ohya, Y. Tanaka, I. Basieva, A. Khrennikov, \emph{Quantum-like model of brain's functioning: decision making from decoherence}, J. Theor. Biol., {\bf 281}, 56-64 (2011)

\bibitem{qdm5} M. Asano, M. Ohya, Y. Tanaka, I. Basieva, A. Khrennikov, \emph{Quantum-like dynamics of decision-making}, Phys. A, {\bf 391}, 2083-2099 (2012)


\bibitem{bagbook} F. Bagarello, \emph{Quantum dynamics for classical
systems: with applications of the Number operator}, Wiley Ed., New York,
(2012)


\bibitem{BO14} F. Bagarello, F. Oliveri, {\em Dynamics of closed ecosystems described by operators}, Ecol. Model. {\bf 275}, 89-
99 (2014)

\bibitem{all1} F. Bagarello, \emph{An operator view on alliances in politics}, SIAM J. on Appl. Math. (SIAP), \textbf{75}, (2), 564--584 (2015)

\bibitem{BGO}{F.~Bagarello, F.~Gargano, F.~Oliveri,} {\em A phenomenological operator description of dynamics of crowds: Escape strategies }, Applied Mathematical Modelling, {\bf 39 (8)}, 2276-2294, (2015)


\bibitem{all2} F. Bagarello, E. Haven, {\em First results on applying a non-linear effect formalism to
alliances between political parties and buy and sell dynamics}, Physica A,  {\bf 444}, 403-414 (2016)


\bibitem{all4}  Bagarello, {\em An improved model of alliances between political parties}, Ricerche di Matematica, doi 10.1007/s11587-016-0261-4, 1-14 (2016)

\bibitem{BCO16} F. Bagarello, A.M. Cherubini, F. Oliveri, {\em An Operatorial Description of Desertification}, SIAM J. Appl. Math., 76(2), 479-499 (2016)



\bibitem{BDGO} F. Bagarello, R. Di Salvo,  F. Gargano, F. Oliveri, {\em $(H,\rho)$-induced dynamics and the quantum game of life},  Applied Mathematical Modelling, \textbf{43 (1)}, 15-32 (2017)



\bibitem{pol3} B. Buonomo, A. d'Onofrio,
 {\em Modeling the Influence of Publics Memory on the Corruption Popularity Dilemma in Politics}, J. Optim. Theory Appl., {\bf 158}, 554-575 (2013)



\bibitem{qdm3} J. R. Busemeyer, P. D. Bruza, \emph{Quantum models of cognition and decision}, Cambridge University Press, Cambridge (2012)

\bibitem{castellano} C. Castellano, S. Fortunato, V. Loreto, {\em Statistical physics of social dynamics}, Rev. Mod. Phys., {\bf 81}, 591-646 (2009)

\bibitem{CCC06} B.K. Chakrabarti, A.Chakraborti, A.Chatterjee, \emph{Econophysics and Sociophysics: Trends and Perspectives}, John Wiley \& Sons (2006).



\bibitem{DO16} R. Di Salvo, F. Oliveri, {\em An operatorial model for long-term survival of bacterial populations}, Ricerche di Matematica, doi:10.1007/s11587-016-0266-z, 1-13 (2016)



\bibitem{FW94} G. Feichtinger, F. Wirl, {\em  On the stability and potential cyclicity of corruption in governments subject to popularity constraints}, Math. Soc. Sci. 28, 113-131 (1994)


\bibitem{galam1} S. Galam, {\em Majority Rule, Hierarchical Structures, and
Democratic Totalitarianism: A Statistical Approach}, Journal of Mathematical
Psychology, \textbf{30}, 426-434 (1986)


\bibitem{pol5}  S. Galam, {\em Sociophysics, A Physicist's Modeling of Psycho-political Phenomena}, Springer (2012)


\bibitem{galam2} S. Galam, \emph{The Drastic Outcomes from Voting Alliances
in Three-Party Democratic Voting (1990$\rightarrow$2013)}, Journal of
Statistical Physics, \textbf{151}, 46-68 (2013)


\bibitem{Gar} F. Gargano, {\em Dynamics of Confined Crowd Modelled Using Fermionic Operators}, Int.~J.~Theor.~Phys., {\bf 53(8)}, 2727-2738, (2014)


\bibitem{qdm2} E. Haven, A. Khrennikov, \emph{Quantum social science},
Cambridge University Press, New York (2013)







\bibitem{qdm1} A. Khrennikov, {\em Ubiquitous quantum structure: from psychology to finances}, Springer,
Berlin, 2010.

\bibitem{pol1} P. Khrennikova, E. Haven, A. Khrennikov, {\em An application of the theory of open quantum systems to model the dynamics of party governance in the US Political System}, Int. Jour. of Theor. Phys., doi: DOI 10.1007/s10773-013-1931-6. (2013)



\bibitem{pol4}  M. Makowski, E. W Piotrowski, {\em Decisions in elections transitive or intransitive quantum preferences},  J. of Phys. A: Mathematical and Theoretical, {\bf 44}, N. 21, 215303 (2011)

\bibitem{mantegna} R.M. Mantegna, E. Stanley, {\em Introduction to
Econophysics}, Cambridge University Press, (1999)


\bibitem{mer} E. Merzbacher. {Quantum Mechanics}, Wiley, New York (1970)

\bibitem{PS07} S.E Page, L.M. Sander, C.M Schneider-Mizell,\emph{Conformity and Dissonance in Generalized Voter Models},
J. Stat. Phys. 128, 1279-1287 (2007)


\bibitem{PT14} F. Palombi, S. Toti,\emph{Stochastic Dynamics of the Multi-State Voter Model Over a Network Based on Interacting Cliques and Zealot Candidates},
J. Stat. Phys. 156, 336-367 (2014)

\bibitem{RM05} G. Raffaelli, M. Marsili,\emph{Statistical mechanics model for the emergence of consensus},
Phys. Rev. E 72, 016114 (2005)

\bibitem{rom} P. Roman, {Advanced quantum mechanics}, Addison--Wesley, New
York (1965)


\bibitem{ST06} D.Stauffer, S.M.M de Oliveira, P.M.C de Oliveira, J.S. de Sà Martins,
\textit{Biology, Sociology, Geology by Computational Physicists
}, Monograph Series on Nonlinear Science and Complexity, Elsevier (2006)







\end{thebibliography}
\end{document}